\documentclass[lineno]{JFM-FLM_Au}

\usepackage{CJKutf8}
\usepackage[dvipsnames]{xcolor} 
\usepackage{bm}

\usepackage{hyperref}
\newcounter{aqctr}
\newenvironment{author-query}
{\refstepcounter{aqctr}\par\vspace{\baselineskip}\noindent
\color{red}\textbf{Author Query/Comment AQ \arabic{aqctr}.}}
{\par\vspace{\baselineskip}\normalcolor}

\lefttitle{I. Kikuchi, H. Watanabe, Y. Yokoyama, H. Kusuno and Y. Tagawa.}
\righttitle{Interfacial dynamics induced by impacts across rigid and soft substrates}

\title{Interfacial dynamics induced by impacts across rigid and soft substrates}

\author{Ishin Kikuchi\aff{1}, Hiroya Watanabe\aff{1}, Yuto Yokoyama\aff{2}, Hiroaki Kusuno\aff{3} \and Yoshiyuki Tagawa\aff{4}}

\affiliation{\aff{1}Department of Mechanical Systems Engineering, Tokyo University of Agriculture and Technology, Tokyo 184-8588, Japan
\aff{2}Micro/Bio/Nanofluidics Unit, Okinawa Institute of Science and Technology, Okinawa 904-0497, Japan
\aff{3}Department of Mechanical Engineering, Kansai University, Osaka 564-8680, Japan
\aff{4}Institute of Global Innovation Research, Tokyo University of Agriculture and Technology, Tokyo 184-8588, Japan}

\corresau{Yoshiyuki Tagawa, \href{tagawayo@cc.tuat.ac.jp}{tagawayo@cc.tuat.ac.jp}}

\begin{document}
\maketitle

\begin{abstract}
We investigate impact-induced gas--liquid interfacial dynamics through experiments in which a liquid-filled container impacts substrates with elastic moduli from $O(10^{-1})$~MPa to $O(10^{5})$~MPa. 
Upon impact, the concave gas--liquid interface inside the container deforms and emits a focused jet.
When the jet velocity is normalized by the container impact velocity, all data collapse onto a single curve when plotted against the Cauchy number, $Ca = \rho_{\rm e} V_{\rm i}^2 / E$, which represents the ratio of the inertial force of the container--liquid system to the elastic restoring force of the substrate.
The dimensionless jet velocity remains nearly constant for $Ca< 10^{-4}$, but decreases significantly for $Ca > 10^{-4}$.
Based on this observation, we define the boundary between the \emph{rigid-impact} 
and \emph{soft-impact} regimes using the Cauchy number, providing a quantitative 
criterion for what constitutes ``softness'' in impact-driven interfacial flows.
To explain the reduction in jet velocity observed in the \emph{soft-impact}  regime, we introduce a framework in which only the impulse transferred within the effective time window for jet formation contributes to interface acceleration.
This concept, referred to as the \emph{partial impulse}, captures the situation where the impact interval (the duration of contact between the container and the substrate) exceeds the focusing interval (the time required for jet formation).
By modelling the contact force using an elastic foundation model and solving the resulting momentum equation over the finite impulse window, we quantitatively reproduce the experimental results.
This \emph{partial impulse} framework unifies the dynamics of impact-driven jetting across both rigid and soft substrate regimes, extending the applicability of classical impulse-based models.
\end{abstract}



\section{Introduction}
\label{sec:introduction}


Impacts are relatively short-duration and violent phenomena involving solids or fluids that occur in a diverse range of situations.
Familiar examples include wave impact on a seawall \citep{cookerPressureimpulseTheoryLiquid1995, peregrineWATERWAVEIMPACTWALLS2003}, skipping stones on a water surface \citep{bocquetPhysicsStoneSkipping2003, roselliniSkippingStones2005,tsaiExperimentalResultsMathematical2022}, and basilisk lizards running on water \citep{glasheenHydrodynamicModelLocomotion1996, glasheenVerticalWaterEntry1996, hsiehRunningWaterThreedimensional2004}.
Engineering cases include violent wave impacts on ships, known as ship slamming \citep{kapsenbergSlammingShipsWhere2011, diasSlammingRecentProgress2018}, strong liquid impacts on container walls due to vibration, called liquid sloshing \citep{cookerPressureimpulseTheoryLiquid1995, peregrineWATERWAVEIMPACTWALLS2003}, the entry of solid objects into liquid \citep{glasheenHydrodynamicModelLocomotion1996, thoroddsenImpactJettingSolid2004, grumstrupCavityRipplesObserved2007, truscottWaterEntryProjectiles2014a, rabbiAlteredDeepsealWater2024}, and the impact of a liquid drop on a solid surface \citep{cookerPressureimpulseTheoryLiquid1995, josserandDropImpactSolid2016a, philippiDropImpactSolid2016, josserandDropletImpactThin2016, philippiPressureImpulseTheory2018}.
When modelling phenomena involving such short-time dynamics, it is extremely effective to consider the impulse or the pressure impulse acting during the impact \citep{cookerPressureimpulseTheoryLiquid1995, glasheenHydrodynamicModelLocomotion1996, glasheenVerticalWaterEntry1996, peregrineWATERWAVEIMPACTWALLS2003,diasSlammingRecentProgress2018, philippiPressureImpulseTheory2018}.

A common feature of impact problems is that they can be defined as the considerable acceleration of a boundary of the system over a short time \citep{philippiPressureImpulseTheory2018}.
One example of an impact problem that exhibits this feature is ``Pokrovski's experiment'' \citep{ lavrentievEffetsHydrodynamiquesModeles1980}.
This experiment is known for the formation of a focused liquid jet from the gas--liquid interface due to the rapid acceleration of the liquid caused by the impact \citep{milgramMotionFluidCylindrical1969,antkowiakShorttermDynamicsDensity2007}.
The details of the formation process of this jet are as follows: when a liquid-filled test tube falls freely, the gas--liquid interface takes on a concave shape due to surface tension in the gravity-free reference frame.
After the tube impacts a rigid substrate, the liquid is rapidly accelerated, and a focused jet is generated from the interface.
In contrast, if the gas--liquid interface is flat just before impact, it does not deform after impact. The liquid simply rebounds together with the container from the substrate, and no jet is produced \citep{antkowiakShorttermDynamicsDensity2007}.

Various studies have investigated Pokrovski's experiment using rigid materials.
\citet{antkowiakShorttermDynamicsDensity2007} pointed out that the acoustic timescale of pressure wave propagation is much shorter than the impact duration, so liquid compressibility can be neglected.
Therefore, the liquid motion during impact can be treated as an incompressible, momentum balance problem—much like a rigid-body impact—and the resulting pressure and velocity fields can be described using pressure--impulse theory.
\citet{kiyama2014} proposed a semi-empirical formula for the jet velocity based on pressure--impulse theory and the focusing effect of a concave gas--liquid interface \citep{tagawaHighlyFocusedSupersonic2012, petersHighlyFocusedSupersonic2013}. 
This formula accurately describes the jet velocity observed in Pokrovski's experiment with a rigid substrate.
Under stronger impacts, cavitation occurs inside the tube, which further increases the jet velocity \citep{kiyamaEffectsWaterHammer2016}.
\citet{watanabeEffectConvergingContainer2025} computed pressure--impulse fields by numerically solving the Laplace equation, using boundary conditions that prescribe the pressure impulse at the container bottom.
They showed that the jet in Pokrovski's experiment is driven by the pressure--impulse gradient.
They also compared the numerically obtained pressure--impulse fields for containers with different convergence angles to analytical solutions from a reduced-order model.
This comparison showed that the jet velocity is controlled by variations in the flow rate and by the pressure--impulse gradient along the centreline, both of which are influenced by changes in the container's cross-sectional area.
\citet{onukiMicrojetGeneratorHighly2018} improved Pokrovski's experiment by inserting a capillary tube into the test tube so that the liquid level inside the capillary was lower than that outside.
This level difference increased the pressure--impulse gradient in the capillary, enabling the ejection of a highly viscous liquid with a kinematic viscosity of 500 mm$^2$/s or higher.
They further pointed out that the jet formation process observed in Pokrovski's experiments can be understood as having two different timescales: the impact interval and the focusing interval. \citet{onukiMicrojetGeneratorHighly2018} defined the impact interval as the stage in which a pressure--impulse gradient generated by the impact rapidly accelerates the liquid, and the focusing interval as the stage in which the liquid is further accelerated by flow focusing at the concave gas--liquid interface after the impact.
By applying high-viscosity jetting devices, impact-based coating systems for viscous liquids have also been developed. \citet{kamamotoDropondemandPaintingHighly2021} succeeded in coating undiluted, highly viscous automotive paint, and \citet{kobayashi2024} achieved continuous coating of silicone oil with a kinematic viscosity of 500 mm$^2$/s at a jetting frequency of 10 Hz.
Other studies have also focused on jet formation in Pokrovski's experiment. 
\citet{chengViscousInfluencesImpulsively2024} investigated the effect of viscosity on jet velocity, while \citet{krishnanImpactFreelyFalling2022} examined the influence of the meniscus shape.
Additionally, \citet{krishnanImpactFreelyFalling2022} and \citet{zhangVelocityScalingBreakup2020} reported that rapid deceleration of the container can also generate a jet, highlighting that large accelerations applied to the container--liquid system are crucial for jet formation.

The insights from Pokrovski's experiment extend beyond jet formation.
For example, \citet{andradeSwirlingFluidReduces2023} and \citet{xiePreliminarySemianalyticalInvestigation2025} conducted experiments similar to Pokrovski's, using containers partially filled with swirling fluid, and reported that fluid dynamics can significantly reduce the rebound of such containers.
They pointed out that the phenomenon can be explained by momentum transfer arising from water redistribution during impact.
Additionally, \citet{panCavitationOnsetCaused2017} focused on cavitation caused by rapid fluid acceleration and proposed a universal criterion to define its onset conditions.

The above reviews Pokrovski's experiment using rigid substrates, in which jet velocities were analysed mainly through pressure--impulse theory under the assumption of incompressibility.
However, Pokrovski's experiment has also been conducted using soft substrates.
\citet{kuriharaPressureFluctuationsLiquids2025} varied the contact time between the container and the substrate from 0.11 ms to 2.2 ms by using metal, resin, and rubber substrates.
They then investigated how the acceleration time of a liquid $\Delta t$ affects the development of the pressure field within the liquid.
They showed that pressure fluctuations in a liquid under rapid acceleration can be characterized by the Strouhal number ($St = L/(c\,\Delta t)$), where $L$ is the liquid filling height and $c$ is the speed of sound), and that incompressible flow theory is applicable when $St \le 0.2$ \citep{kuriharaPressureFluctuationsLiquids2025}.
They also proposed a conceptual model that introduces a modified water-hammer theory accounting for the finite thickness of the pressure wavefront.
In a related study, \citet{micheleWeaklyNonlinearTheory2025} discussed the water-hammer phenomenon that occurs when a valve is operated slowly.
They solved the nonlinear governing equations using a perturbation scheme combined with the Laplace transform, obtaining analytical solutions that show how the flow changes in space and time throughout the pipe.
The study by \citet{micheleWeaklyNonlinearTheory2025} is closely related to that of \citet{kuriharaPressureFluctuationsLiquids2025} in that it considers a finite liquid acceleration time by slowly operating a valve.

In the following, we organize the existing findings on Pokrovski's experiments and then present the objective of this study.
In this study, the impact interval is redefined as the duration for which the container and the substrate remain in contact, while the focusing interval is redefined as the time required for a jet to form from a concave gas--liquid interface.
Under this interpretation, the study by \citet{kuriharaPressureFluctuationsLiquids2025} can be understood as elucidating the role of the liquid acceleration time in the formation of the pressure field during the impact interval.
However, \citet{kuriharaPressureFluctuationsLiquids2025} did not discuss the effect of a soft substrate on the focusing interval.

Here we extend Pokrovski's experiment to compliant substrates in order to isolate how substrate deformability alters the impulse used to deform the concave gas--liquid interface (hereafter denoted as \textit{I}) and, ultimately, the velocity of the focused jet.
By varying the substrate elastic modulus over nearly six orders of magnitude (from $8.1 \times 10^{-1}$ MPa to $2.0 \times 10^{5}$ MPa), we lengthen the contact time from sub-milliseconds to 8 ms and access a regime in which the contact interval (i.e. the impact interval) overlaps with the jet-formation interval (i.e. the focusing interval).
In this regime, the jet forms while the container is still in contact with the substrate, implying that the liquid interface cannot be driven by the fully transmitted impulse (i.e. the \emph{total impulse}) assumed in conventional rigid substrate descriptions.

Our important result is that the dimensionless jet velocity collapses when plotted against the Cauchy number $\rho_{\rm e} V_{\rm i}^2 / E$, and decreases markedly for $\rho_{\rm e} V_{\rm i}^2 / E > 10^{-4}$.
We interpret this decrease as a \emph{partial impulse} effect: when the impact interval exceeds the focusing interval, only the impulse delivered within an effective time window set by the focusing dynamics contributes to jet acceleration.
We formalize this idea by introducing an effective impulse window and combining it with an elastic foundation model, which together provide a quantitative description of jet velocities across both rigid and compliant substrates.

This paper is organized as follows.
\S \ref{sec:experimental setup} describes the experimental setup and the data analysis.
\S \ref{sec:experimental observations} presents the observed dependence of the jetting dynamics on the substrate stiffness.
\S \ref{sec:scaling law} shows that the data collapse when scaled by $\rho_{\rm e} V_{\rm i}^2 / E$ and identifies a compliant impact regime for $\rho_{\rm e} V_{\rm i}^2 / E > 10^{-4}$, where the conventional rigid substrate picture fails.
\S \ref{sec:mechanism} develops the \emph{partial impulse} framework and a predictive model, and compares the model with the experimental results.
\S \ref{sec:conclusion} summarizes the main conclusions.

\section{Experimental setup}
\label{sec:experimental setup}

\begin{figure}
  \centerline{\includegraphics[width=0.65\textwidth]{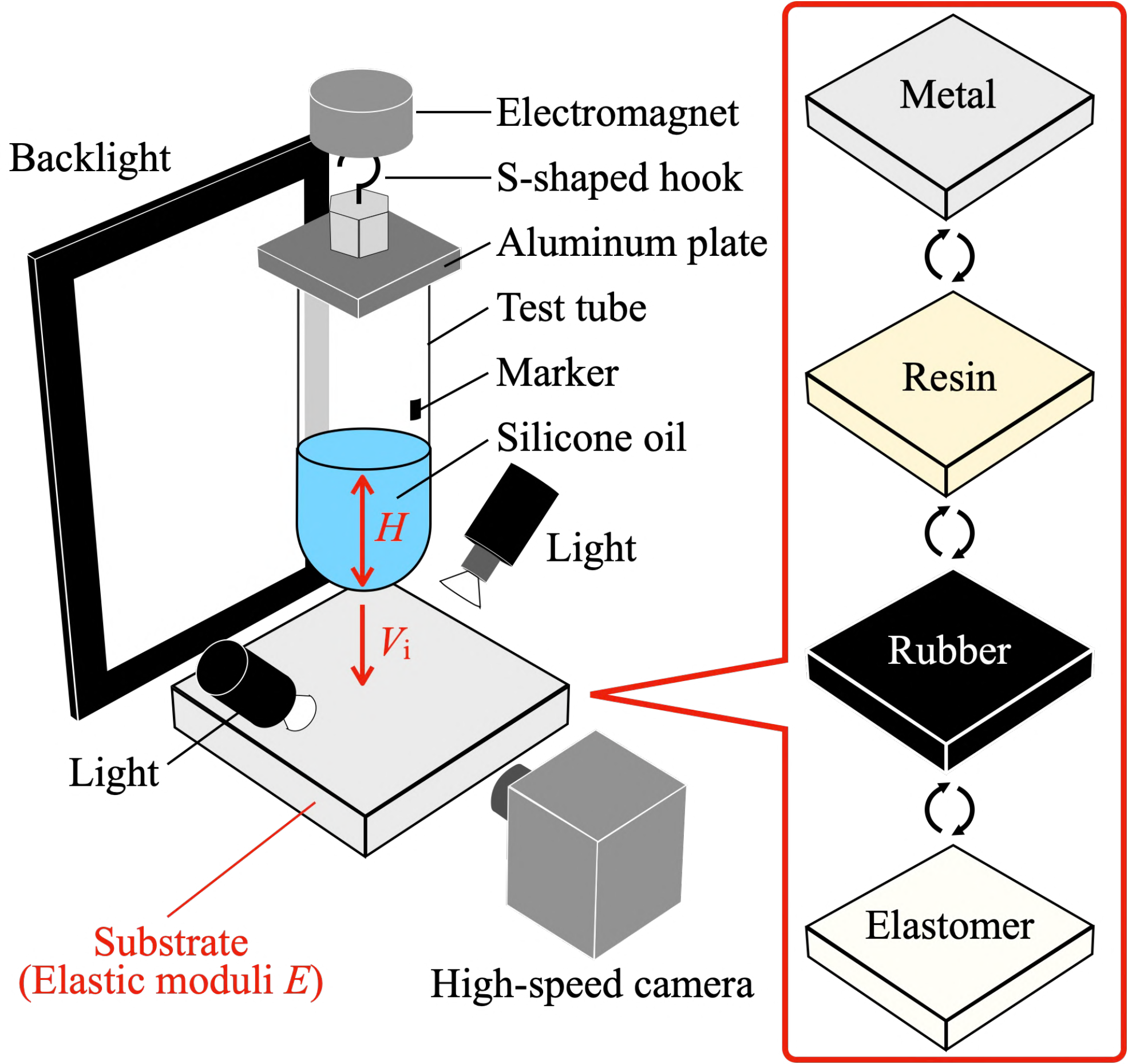}}
  \captionsetup{style=capcenter}
  \caption{A schematic of the experimental setup. A test tube filled with silicone oil of 10 mm$^2$/s was released in free fall using an electromagnet to impact the substrate. The experimental parameters were the elastic moduli of the substrates $E$, the impact velocity of the impactor $V_{\rm{i}}$, and the liquid filling height $H$. Nine substrates with different $E$ made of metals, resins, rubber and elastomers were used.}
\label{fig:exp}
\end{figure}

To investigate how a soft substrate affects the impulse $I$ used to deform a concave gas--liquid interface, we conducted Pokrovski's experiment using nine substrates with different elastic moduli.
Figure \ref{fig:exp} shows the experimental setup.
First, a test tube (Maruemu, A-16.5, inner diameter 14.3 mm, outer diameter 16.5 mm) was filled with silicone oil of 10 mm$^2$/s (Shin-Etsu Chemical, KF-96-10CS).
Next, an aluminium plate with an iron S-shaped hook was attached to the top of the test tube using adhesive.
The test tube was then placed on the substrate using an electromagnet (Fujita, round electromagnet FSGP-40).
When the electromagnet was turned off, the test tube fell freely, and the gas--liquid interface was deformed by surface tension into a concave shape.
After impact with the substrate, the interface deformed significantly, and a focused liquid jet was ejected.
Two lights (HAYASHI-REPIC, LA-HDF158AA) were placed above the substrate to make it easier to observe the contact between the container and the substrate.

Table \ref{table} summarizes the substrate material, elastic modulus $E$, size and thickness $h$, as well as the experimental conditions: the liquid filling height $H$ and the impact velocity $V_{\rm{i}}$ of the impactor, which consisted of the test tube, the liquid, the aluminium plate and the S-shaped hook as a single unit.
The experimental parameters were $V_{\rm{i}}$, $H$ and $E$.
$V_{\rm i}$ was varied from 0.63 m/s to 1.1 m/s by changing the drop height $h_{\rm drop}$ from 20 mm to 60 mm.
Here, $V_{\rm i}$ was calculated using $V_{\rm i} = \sqrt{2gh_{\rm drop}}$, where $g = 9.81$ m/s$^2$ is the gravitational acceleration.
$H$ was used to change the mass of the impactor.
For a rigid substrate, \citet{kiyamaEffectsWaterHammer2016} reported that the pressure--impulse gradient generated in the liquid at impact does not depend on $H$.
Therefore, except when cavitation occurs, the jet velocity is independent of $H$.
In contrast, in this study, we examine the effect of $H$ on the jet velocity when $E$ is small.
For the metal substrate, $H$ was set to 20 mm and 40 mm, while for the softer substrates with lower elasticity, $H$ was set to 20 mm, 40 mm and 60 mm.
For all settings for $H$, $V_{\rm i}$ and $E$, we checked the high-speed images to confirm whether cavitation occurred, and no cavitation larger than one pixel was observed.
Therefore, we concluded that no cavitation contributing to jet acceleration, as reported by \citet{kiyamaEffectsWaterHammer2016}, occurred in our experiments.
$E$ was varied from $8.1 \times 10^{-1}$ MPa to $2.0 \times 10^{5}$ MPa by using nine types of substrates made of metal, resin, rubber and elastomer.
For metal and resin substrates, the values of $E$ were taken from the literature \citep{ashbyEngineeringMaterials12012, yahamedMechanicalProperties3D2016}.
The $E$ values of the rubber and elastomer substrates were determined experimentally by pressing a rigid sphere against the substrate surface \citep{yokoyamaHighspeedPhotoelasticTomography2024}.
In this experiment, the load applied to the substrate was measured as a function of the imposed surface displacement.
$E$ was then obtained by applying an elastic foundation model that describes the relationship between the substrate displacement and the applied load \citep{johnsonContactMechanics1987} (see Appendix \ref{app:measuring E}).
We fabricated the elastomer substrates using polydimethylsiloxane (PDMS) (see Appendix \ref{app:PDMS}).
The elastic modulus of PDMS can be adjusted by changing the mass ratio of the base polymer (DOW, SILPOT 184 Silicone Elastomer Base) to the curing agent (DOW, SILPOT 184 Silicone Elastomer Curing Agent) \citep{shuklaSubstrateStiffnessModulates2016}.
We used three types of PDMS substrates with mass ratios of 10:1 (PDMS10), 20:1 (PDMS20) and 30:1 (PDMS30).
To remove the stickiness of the PDMS substrate, baby powder (SiCCAROL-Hi, Asahi Group Foods Co., Ltd.) was applied to eliminate surface adhesion.

Images were captured using a high-speed camera (Photron, FASTCAM SA-X) with backlighting (PHLOX, White LED Backlight $400 \times 200$ mm)
The frame rate of the camera was 10 000 fps, and the resolution was 0.160--0.169 mm/pixel.
The images were used to analyse the jet velocity, the time at which contact between the container and the substrate ended ($\tau_{\rm impact}$), the time at which the centre of the interface rose due to flow focusing and the protrusion became visible from the side ($\tau_{\rm focusing}$), and the gas--liquid interface thickness $H_{\rm m}$ (as discussed in \S \ref{sec:results and discussion}).
$\tau_{\rm focusing}$ can be visually identified from the consecutive jet images in figure \ref{fig:ejection}.
The jet velocity was calculated from the relative position of the jet tip with respect to the test tube.
The position of the test tube relative to the substrate was calculated using the printed marker on the test tube.
For each experimental condition, the representative jet velocity $V_{\rm j}$ was defined as the jet velocity at $t=\tau_{\rm focusing}$ (see \S\ \ref{sec:experimental observations}).
From the high-speed camera images, it was confirmed that the concave shape of the gas--liquid interface just before impact, which strongly affects the motion of the gas--liquid interface \citep{antkowiakShorttermDynamicsDensity2007}, did not change significantly across all conditions.
The jet generation was repeated five times for each set of conditions.


\begin{table}
  \begin{center}
\def~{\hphantom{0}}
  \begin{tabular}{ccccccc}

Substrate & Material & $E$ {[}MPa{]} & Size {[}mm{]} & $h$ {[}mm{]} & $H$ {[}mm{]} & $V_{\rm{i}}$ {[}m/s{]}\\

SS400 & Steel & $\bm{2.0 \times 10^{5}}$ & $150 \times 150$ & 20 & 20-40 & 0.63-1.1\\
A5052 & Aluminium & $\bm{6.9 \times 10^{4}}$ & $150 \times 150$ & 20 & 20-40 & 0.63-1.1\\
Epoxy & Epoxy resin & $\bm{3.0 \times 10^{3}}$ & $150 \times 150$ & 20 & 20-60 & 0.63-1.1\\
ABS &Acrylonitrile Butadiene Styrene resin & $\bm{1.9 \times 10^{3}}$ & $150 \times 150$ & 20 & 20-60 & 0.63-1.1\\
PE & Polyethylene resin & $\bm{7.0 \times 10^{2}}$ & $150 \times 150$ & 20 & 20-60 & 0.63-1.1\\
Rubber & Rubber & $\bm{1.9 \times 10^{1}}$ & $150 \times 150$ & 22.5 & 20-60 & 0.63-1.1\\
PDMS10 & Polydimethylsiloxane & $\bm{5.7 \times 10^{0}}$ & $100 \times 100$ & 22.0 & 20-60 & 0.63-1.1\\
PDMS20 & Polydimethylsiloxane & $\bm{1.7 \times 10^{0}}$ & $100 \times 100$ & 22.5 & 20-60 & 0.63-1.1\\
PDMS30 & Polydimethylsiloxane & $\bm{8.1 \times 10^{-1}}$ & $100 \times 100$ & 22.0 & 20-60& 0.63-1.1

  \end{tabular}
  \caption{A summary of the substrate materials, elastic moduli $E$, sizes, and thicknesses $h$, together with the values of the liquid filling height $H$ and impactor impact velocity $V_{\rm i}$ for each substrate.}
  \label{table}
  \end{center}
\end{table}

\section{Results and discussion}
\label{sec:results and discussion}

In this section, \S \ref{sec:experimental observations} describes the typical interface motion and jet behaviour induced by impact onto substrates with various $E$.
We also discuss the relationship between the substrate elasticity and the jet velocity.
In \S \ref{sec:scaling law}, we show that the jet velocity can be described in a unified manner using the Cauchy number by organizing the experimental results through dimensional analysis.
In \S \ref{sec:mechanism}, we clarify the mechanism of the reduced jet velocity on substrates with small $E$ by discussing the impulse $I$ used to deform a concave gas--liquid interface with proposing a concept of a \emph{partial impulse}.

\subsection{Experimental observations of jet velocity reduction on soft substrates}
\label{sec:experimental observations}

\begin{figure}
  \centerline{\includegraphics[width=1\textwidth]{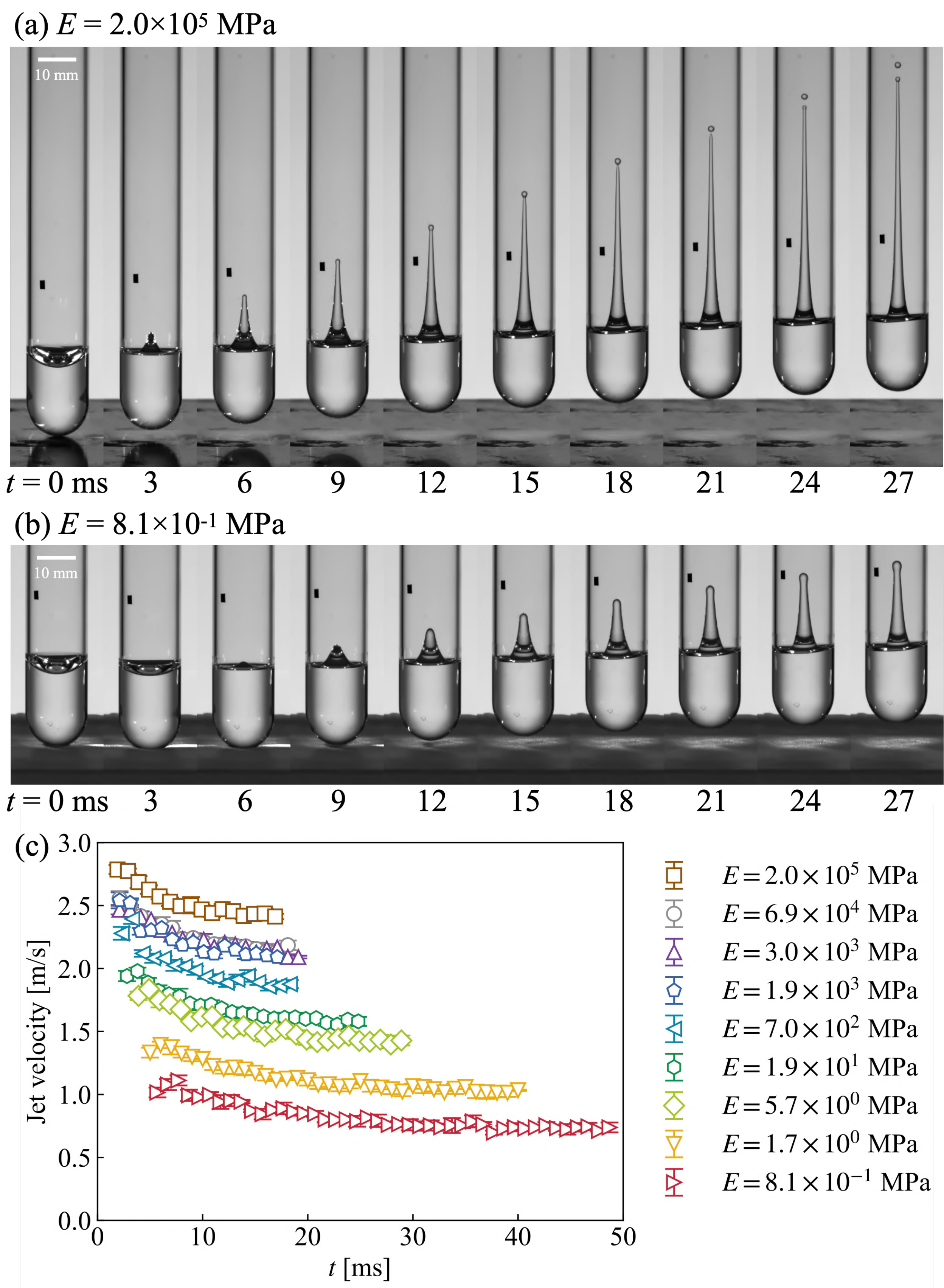}}
  \captionsetup{style=capcenter}
  \caption{A comparison of the interface motion and jet behaviour for $H = 20$ mm and $V_{\rm i} = 0.63$ m/s: (a) $E = 2.0 \times 10^{5}$ MPa, (b) $E = 8.1 \times 10^{-1}$ MPa. $t = 0$ ms represents the time at which the container first makes contact with the substrate. For $E = 8.1 \times 10^{-1}$ MPa, the time required for the concave interface to deform is longer, and the jet tip position is lower, compared with $E = 2.0 \times 10^{5}$ MPa. (c) The time evolution of the jet velocity for different values of $E$. Here, $t = 0$ ms represents the time at which the container first makes contact with the substrate; the plot starts at $t=\tau_{\rm focusing}$ and ends at the time of pinch-off. The results indicate that as $E$ decreases, the jet velocity becomes slower, while $\tau_{\rm focusing}$ increases.}
\label{fig:img & Vtip}
\end{figure}

In this section, we show that the jet velocity decreases as the substrate elasticity decreases.
We also show that, for softer substrates, the time required to deform the concave gas--liquid interface clearly increases.
Figures \ref{fig:img & Vtip}(a) and \ref{fig:img & Vtip}(b) show the interface motion and the jet behaviour for $E = 2.0 \times 10^{5}$ MPa and $E = 8.1 \times 10^{-1}$ MPa at $H = 20$ mm and $V_{\rm i} = 0.63$ m/s.
The time at which the container first makes contact with the substrate is defined as $t = 0$ ms, and the time interval between each image is 3 ms.
As mentioned in \S \ref{sec:experimental setup}, when the electromagnet was turned off, the impactor fell freely and impacted the substrate.
Following the impact, the flow was focused by the concave gas--liquid interface, leading to the ejection and elongation of a focused liquid jet.
At $t = 27$ ms, the jet tip was lower for $E = 8.1 \times 10^{-1}$ MPa than for $E = 2.0 \times 10^{5}$ MPa, which shows that the jet velocity was reduced.
Furthermore, the time required for the concave interface to deform and generate a jet was longer for $E = 8.1 \times 10^{-1}$ MPa than for $E = 2.0 \times 10^{5}$ MPa.
To quantify this difference, we investigated the time $t=\tau_{\rm focusing}$ when flow focusing was completed and the central part of the interface rose, making the protrusion visible from the side.
At $E = 2.0 \times 10^{5}$ MPa, $\tau_{\rm focusing}=1.9$ ms, while at $E = 8.1 \times 10^{-1}$ MPa, $\tau_{\rm focusing}=5.7$ ms, which is about three times longer.

Figure \ref{fig:img & Vtip}(c) shows the time evolution of the jet velocity in 1 ms intervals for each $E$ at $H = 20$ mm and $V_{\rm i} = 0.63$ m/s.
$t = 0$ ms denotes the time at which the container first makes contact with the substrate; the plot starts at $t=\tau_{\rm focusing}$ and ends at the time of pinch-off.
Each plot represents the average of five trials, and the error bars indicate the standard deviation.
It is clearly observed that the jet velocity decreases as $E$ decreases.
Looking at the starting points of the $t=\tau_{\rm focusing}$ plots for each $E$, $\tau_{\rm focusing}$ increases with decreasing $E$.
Moreover, the ending points of the plots also increase as $E$ decreases, showing that the jet growth becomes slower.
Focusing on the time evolution of the jet velocity, it reaches its maximum around $t=\tau_{\rm focusing}$ and then converges to a constant value.
In the following sections, we use the representative jet velocity $V_{\rm j}$, defined as the jet velocity at $t=\tau_{\rm focusing}$.

\begin{figure}
  \centerline{\includegraphics[width=1\textwidth]{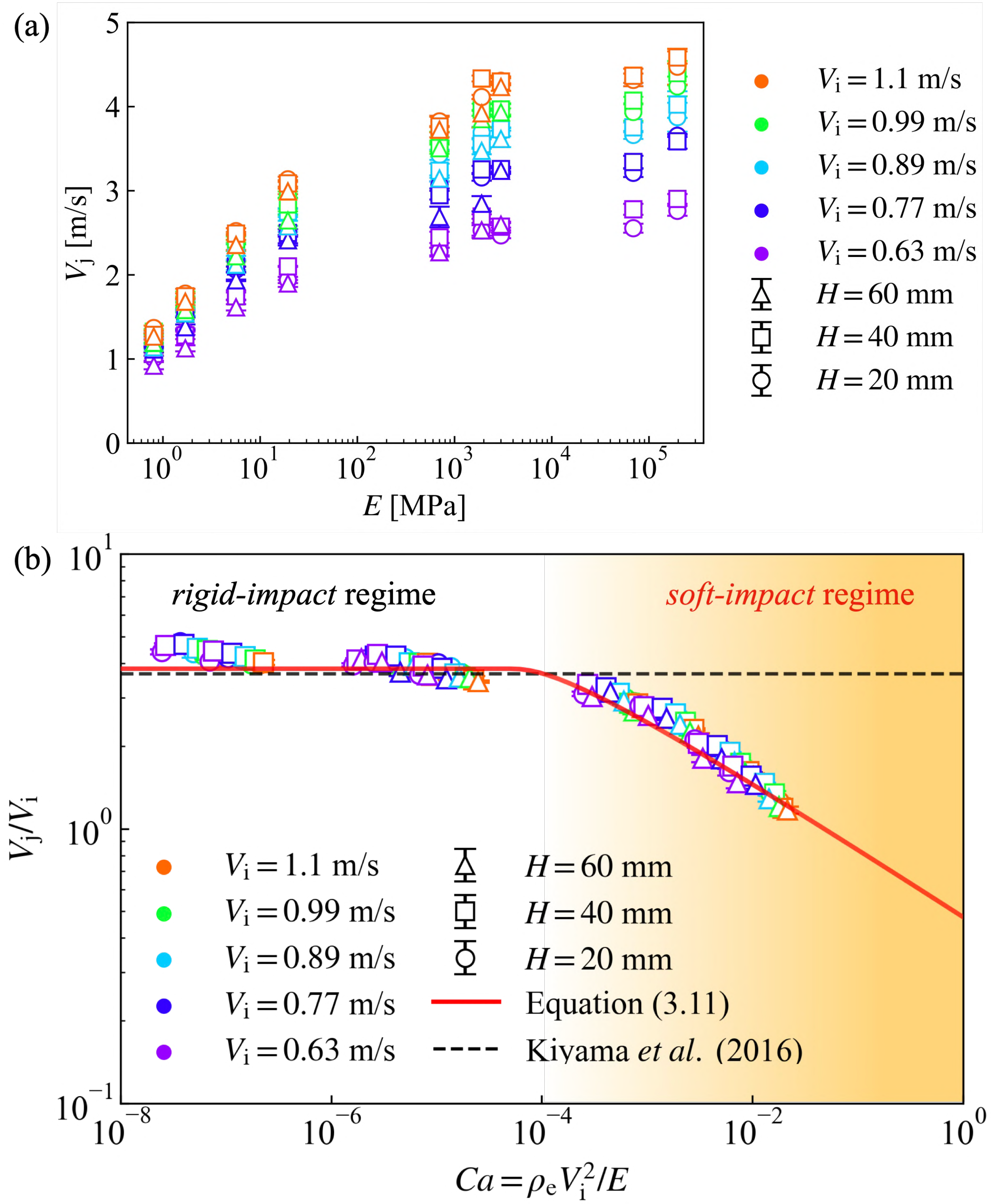}}
  \captionsetup{style=capcenter}
  \caption{(a) $V_{\rm{j}}$ vs. $E$. $V_{\rm j}$ increases with increasing $E$ at fixed $V_{\rm i}$ and also increases with $V_{\rm i}$ at fixed $E$, indicating that $V_{\rm j}$ strongly depends on both $E$ and $V_{\rm i}$. However, the data points do not collapse onto a single curve. (b) $V_{\rm{j}}/V_{\rm{i}}$ vs. $\rho_{\rm{e}}V_{\rm{i}}^2/E$. The black dashed line shows the $V_{\rm j}/V_{\rm i} = 3.66$ estimated from the rigid substrate experiments by \citet{kiyamaEffectsWaterHammer2016}, and the red line shows equation (\ref{eq:Vj model}).
  All data collapse onto a single curve. When $\rho_{\rm e} V_{\rm i}^2 / E< 10^{-4}$, the variation in $V_{\rm j}/V_{\rm i}$ is small.
Since this regime can be explained by the pressure--impulse theory established in previous studies using rigid substrates, we refer to it as the \emph{rigid-impact}  regime in this paper.
In contrast, when $\rho_{\rm e} V_{\rm i}^2 / E > 10^{-4}$, $V_{\rm j}/V_{\rm i}$ clearly decreases.
This regime cannot be explained by conventional pressure--impulse theory. Because the influence of the soft substrate on the interfacial motion is pronounced, we refer to this region as the \emph{soft-impact}  regime.}
\label{fig:Vj & scaling}
\end{figure}

Next, we discuss the relationship between $V_{\rm j}$ and $E$.
Figure \ref{fig:Vj & scaling}(a) shows the relationship between $V_{\rm j}$ and $E$ for all experimental conditions on a semi-log plot.
Each plot represents the average of five trials, and the error bars indicate the standard deviation.
Overall, $V_{\rm j}$ tends to increase as $E$ increases, but the data points do not collapse onto a single curve.
In more detail, when $V_{\rm i}$ is fixed, $V_{\rm j}$ increases with increasing $E$, and when $E$ is fixed, $V_{\rm j}$ increases with increasing $V_{\rm i}$.
These results show that $V_{\rm j}$ is strongly affected by both $E$ and $V_{\rm i}$.
In the following sections, we derive a dimensionless number that can describe the relationship between $V_{\rm j}$ and $E$ in a unified manner using dimensional analysis.

\subsection{Jet velocity and Cauchy number: Dimensional analysis}
\label{sec:scaling law}

In this section, we use dimensional analysis to show that the experimental jet velocities can be unified using the Cauchy number.
We also show that for a Cauchy number smaller than approximately $10^{-4}$, the jet velocity can be described by the conventional pressure--impulse theory; this range is defined here as the \emph{rigid-impact}  regime.
Furthermore, for a Cauchy number larger than approximately $10^{-4}$, the jet velocity decreases in a way that cannot be explained by conventional pressure--impulse theory.
In this regime, the influence of the soft substrate on the interfacial motion becomes pronounced; therefore, we define this regime as the \emph{soft-impact}  regime.

We use dimensional analysis to derive a dimensionless number that collapses the relationship between $V_{\rm j}$ and $E$.
Since the bottom of the impactor is hemispherical, the impact can be approximated as an impact between a rigid sphere and the substrate.
We also assume that $V_{\rm j}$ can be expressed as a function of $E$, $V_{\rm i}$, and the effective density of the impactor, $\rho_{\rm e}$, as

\begin{equation}
 V_{\rm j} \propto  E^a V_{\rm i}^b \rho_{\rm e}^c.
  \label{eq:Vj_function}
\end{equation}

\noindent
Here, $a$, $b$, and $c$ are power exponents.
$\rho_{\rm e}$ is calculated from the impactor mass $m$ as follows:

\begin{equation}
\rho_{\rm e} = \frac{3m}{4 \pi R^3},
  \label{eq:rho_e}
\end{equation}

\noindent
which represents the density of a rigid sphere with the same outer diameter, $2R$, as the test tube, and varies with $H$ in this experiment.
Within the range of experimental conditions, $H$ does not significantly affect $V_{\rm j}$ (see figure \ref{fig:Vj & scaling}(a)).
However, when a rigid sphere impacts a soft substrate, the sphere's density is one of the parameters that determines the type of contact, such as elastic or viscoelastic \citep{maruokaFrameworkCrossoverScaling2023}.
Because the contact force on the container from the substrate strongly depends on the contact type and affects the jet velocity, this study considers $\rho_{\rm e}$ as a parameter for $V_{\rm j}$.

Dimensional analysis gives

\begin{equation}
\frac{V_{\rm j}}{V_{\rm i}} \propto   \Big(\frac{\rho_{\rm e} V_{\rm i}^2}{E}\Big)^{-a},
  \label{eq:dimensional analysis 2}
\end{equation}

\noindent
where $V_{\rm j}/V_{\rm i}$ represents the dimensionless jet velocity, obtained by normalizing $V_{\rm j}$ with $V_{\rm i}$.
$\rho_{\rm e} V_{\rm i}^2 / E$ is a dimensionless number known as the Cauchy number, which represents the ratio of the impactor's inertial force to the substrate's elastic force.
The dimensional analysis shows that $V_{\rm j} / V_{\rm i}$ can be scaled by the Cauchy number.

We next investigate the relationship between $V_{\rm j}/V_{\rm i}$ and $\rho_{\rm e}V_{\rm i}^2/E$ in detail using the experimental data.
Figure \ref{fig:Vj & scaling}(b) shows the dependence of $V_{\rm j}/V_{\rm i}$ on $\rho_{\rm e}V_{\rm i}^2/E$.
Each plot represents the average of five trials, and the error bars indicate the standard deviation.
Compared with figure \ref{fig:Vj & scaling}(a), all the plots in figure \ref{fig:Vj & scaling}(b) collapse onto a single curve.
A key feature of the relationship between $V_{\rm j}/V_{\rm i}$ and $\rho_{\rm e}V_{\rm i}^2/E$ is that around $\rho_{\rm e}V_{\rm i}^2/E \sim 10^{-4}$, the data separates into two regions: one where $V_{\rm j}/V_{\rm i}$ changes little, and another where it clearly decreases.

In the region where $\rho_{\rm e}V_{\rm i}^2/E< 10^{-4}$, the value of $V_{\rm j}/V_{\rm i}$ changes only slightly.
This behaviour can be explained using the results of \citet{kiyamaEffectsWaterHammer2016}, who reported the jet velocity in Pokrovski's experiment with metal substrates.
 \citet{kiyamaEffectsWaterHammer2016} reported that the dimensionless jet velocity $V_{\rm j}/U_0$, where $U_0$ is the velocity of the concave gas--liquid interface just after the impact, becomes $V_{\rm j}/U_0 = 2.05$ when cavitation does not occur.
We estimate $V_{\rm j}/V_{\rm i}$ from $V_{\rm j}/U_0 = 2.05$, as reported by \citet{kiyamaEffectsWaterHammer2016}, and compare it with our experimental data.
$U_0$ is defined as $U_0 = V_{\rm i} + V_{\rm r}$, where $V_{\rm r}$ is the rebound velocity of the impactor relative to the substrate \citep{kiyamaEffectsWaterHammer2016}.
Because $V_{\rm r}$ can be estimated as $ V_{\rm r} \approx eV_{\rm i}$ using the restitution coefficient $e$, $U_0$ can be approximated by $U_0 \approx V_{\rm i}(1 + e)$.
Therefore, to estimate $V_{\rm j}/V_{\rm i}$ from the value of $V_{\rm j}/U_0$ reported by \citet{kiyamaEffectsWaterHammer2016}, we can use
$V_{\rm j}/V_{\rm i} \approx V_{\rm j}(1+e)/U_0$.
$e$ was obtained by analysing $V_{\rm r}$ for the steel and aluminium substrates used in our experiments, and calculating $e = V_{\rm r}/V_{\rm i}$ to give $e = 0.787$.
We obtained $V_{\rm r}$ using the same method as \citet{kiyamaEffectsWaterHammer2016}.
We estimated $V_{\rm j}/V_{\rm i}$ from $V_{\rm j}/U_0$ reported by \citet{kiyamaEffectsWaterHammer2016}, and obtained $V_{\rm j}/V_{\rm i} = 3.66$.
The black dashed line in figure \ref{fig:Vj & scaling}(b) represents $V_{\rm j}/V_{\rm i} = 3.66$.
We find that the result of \citet{kiyamaEffectsWaterHammer2016} captures the trend of $ V_{\rm j}/V_{\rm i}$ well in the region where $\rho_{\rm e} V_{\rm i}^2 / E< 10^{-4}$ in our experiments.
This indicates that the jet velocity for $\rho_{\rm e} V_{\rm i}^2 / E< 10^{-4}$ can be explained by the theory established from Pokrovski's experiments using metal substrates.
So far, the impact of a “fluid-containing container” on a rigid substrate has been modelled similarly to impacts between rigid bodies, using an incompressible-based approximation in which the \emph{total impulse} from the substrate is transmitted almost instantaneously \citep{batchelorIntroductionFluidDynamics1967, antkowiakShorttermDynamicsDensity2007, kiyama2014, kiyamaEffectsWaterHammer2016}.
The present results show that the applicability of this incompressible-based approximation can be evaluated using the Cauchy number.
Furthermore, it is experimentally confirmed for the first time that the incompressible-based approximation is valid when  $\rho_{\rm e} V_{\rm i}^2 / E< 10^{-4}$.
Therefore, in this study, the region $\rho_{\rm e} V_{\rm i}^2 / E< 10^{-4}$ is defined as the \emph{rigid-impact}  regime.

On the other hand, in the region where $\rho_{\rm e} V_{\rm i}^2 / E > 10^{-4}$, $V_{\rm j}/V_{\rm i}$ decreases significantly with increasing Cauchy number.
This region cannot be modelled using the incompressible-based approximation.
In this study, we experimentally show for the first time that the effect of a small $E$ becomes significant in fluid motion when $\rho_{\rm e} V_{\rm i}^2 / E > 10^{-4}$.
Therefore, we define the region $\rho_{\rm e} V_{\rm i}^2 / E > 10^{-4}$ as the \emph{soft-impact}  regime.

Since the \emph{rigid-impact}  regime and the \emph{soft-impact}  regime can be evaluated using the Cauchy number, these two regimes are determined not only by $E$ but also by the contact conditions, including $V_{\rm i}$ and $\rho_{\rm e}$.
However, under the present experimental conditions, the value of Cauchy number is predominantly determined by $E$.
Therefore, the classification into the \emph{rigid-impact} and \emph{soft-impact} regimes can be readily made based on $E$.
In the present experiments, all data for $E \le 1.9 \times 10^{1}$ MPa are classified into \emph{soft-impact} regime, whereas all data for $E \ge 7.0 \times 10^{2}$ MPa fall within \emph{rigid-impact} regime.
In the next section, we discuss the decrease of $V_{\rm j}/V_{\rm i}$ in the \emph{soft-impact}  regime and clarify its underlying mechanism.
Other studies have also used Cauchy number scaling to describe fluid motion with impact and the deformation of rigid bodies.
For example, \citet{huBreakingWaveImpacts2025} experimentally investigated the hydroelastic response of a vertical cantilever plate subjected to breaking-wave impacts.
They showed that the plate deflection caused by the wave impact can be scaled by the Cauchy number, providing practical insights into hydroelastic effects under various impact conditions.

\subsection{Mechanism of jet velocity reduction in the soft-impact regime}
\label{sec:mechanism}

In this section, we elucidate the physical mechanism leading to the reduction of $V_{\rm j}/V_{\rm i}$ observed for the \emph{soft-impact}  regime.
The key observation is the relative ordering of two timescales: the impact interval $\tau_{\rm impact}$, during which the container and substrate remain in contact, and the focusing interval $\tau_{\rm focusing}$, required for the concave gas--liquid interface to focus and produce a visible jet.
For rigid substrates, $\tau_{\rm impact} \ll \tau_{\rm focusing}$, so the contact force acts over a short interval, and the fluid subsequently focuses after the contact has ended.
In this limit, it is natural to relate jetting to the \emph{total impulse} delivered during contact.

For sufficiently soft substrates, however, we find $\tau_{\rm impact} > \tau_{\rm focusing}$, meaning that jetting starts while contact is ongoing.
In that case, the interface cannot be driven by the \emph{total impulse} transmitted over the entire contact, because the jet forms before that impulse has been fully delivered.
This motivates a framework in which the jet is controlled by the impulse accumulated up to the jet-formation time, i.e., the \emph{partial impulse}.
Below we formalize this idea by (i) writing a momentum balance over the effective time window relevant for jet formation and (ii) modelling the contact force history using an elastic foundation model.

\subsubsection{Relative magnitude between the impact interval and the focusing interval}\label{subsec:impact&focusing interval}

\begin{figure}
  \centerline{\includegraphics[width=1\textwidth]{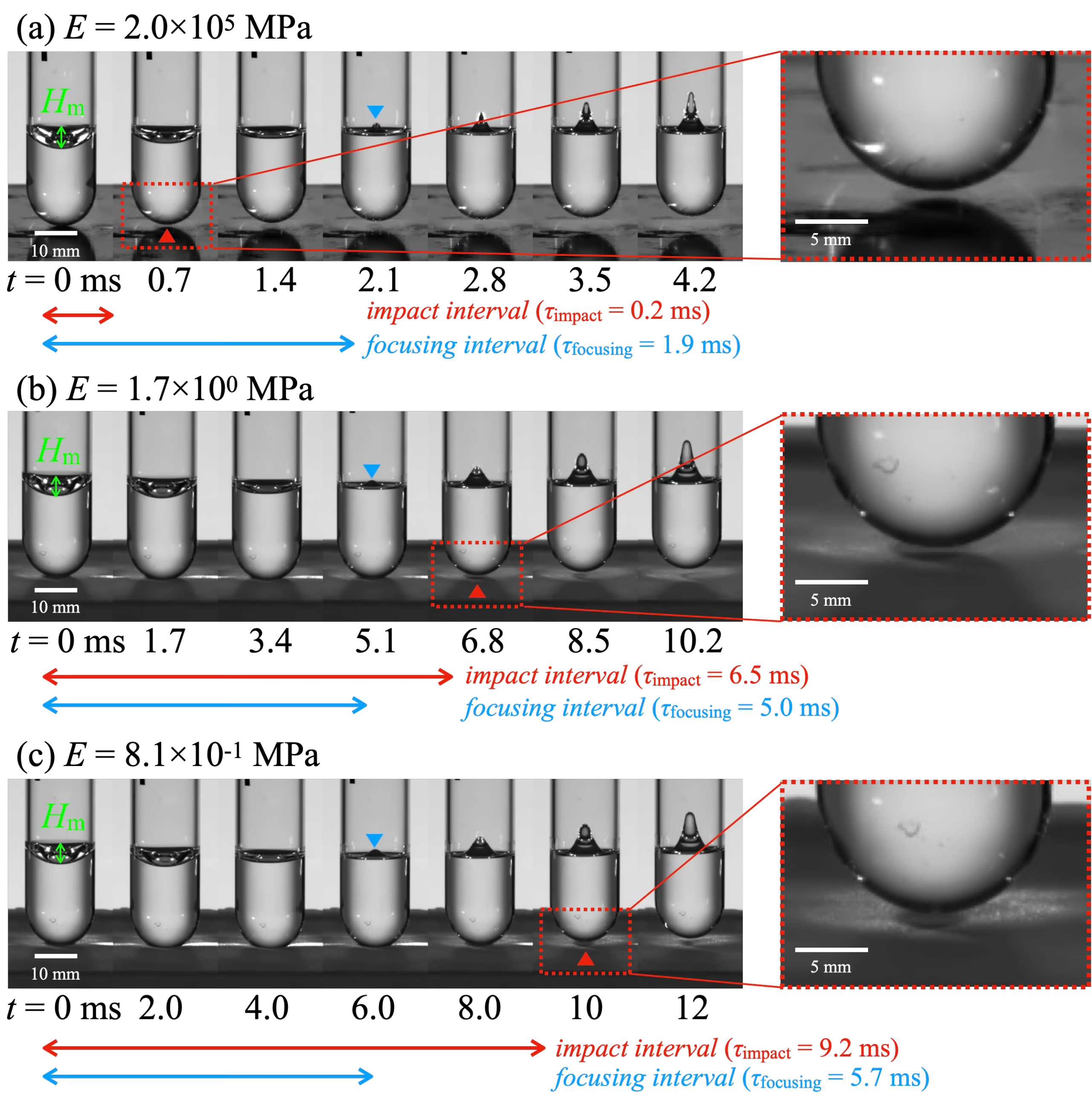}}
  \captionsetup{style=capcenter}
  \caption{The moment when the container separates from the substrate (see \textcolor{red}{$\blacktriangle$}) and when the centre of the interface rises and the protrusion becomes visible from the side (see \textcolor{ProcessBlue}{$\blacktriangledown$}) are shown for (a) $E = 2.0 \times 10^{5}$ MPa, (b) $E = 1.7 \times 10^{0}$ MPa and (c) $E = 8.1 \times 10^{-1}$ MPa ($H = 20$ mm and $ V_{\rm i} = 0.63$ m/s). 
$t = 0$ ms denotes the time when the container begins to make contact with the substrate.
For $E = 2.0 \times 10^{5}$ MPa, $\tau_{\rm impact} = 0.2$ ms and $\tau_{\rm focusing} = 1.9$ ms, giving $\tau_{\rm impact} \ll \tau_{\rm focusing}$. In other words, the impact interval is short compared to the focusing interval. In contrast, for $E = 1.7 \times 10^{0}$ MPa and $E = 8.1 \times 10^{-1}$ MPa, the centre of the interface rises, and the protrusion becomes visible from the side while the container and the substrate are in contact. In other words, the impact interval becomes longer than the focusing interval. $H_{\rm m}$ represents the meniscus thickness just before the container and the substrate come into contact.}
\label{fig:ejection}
\end{figure}

Figure \ref{fig:ejection} shows the moment when the container separates from the substrate (see \textcolor{red}{$\blacktriangle$}) and the moment when the centre of the interface rises and a protrusion becomes visible from the side (see \textcolor{ProcessBlue}{$\blacktriangledown$}) for $H = 20$ mm and $V_{\rm i} = 0.63$ m/s.
Panels (a), (b) and (c) correspond to $ E = 2.0 \times 10^{5}$ MPa, $E = 1.7 \times 10^{0}$ MPa and $E = 8.1 \times 10^{-1}$ MPa, respectively.
The time $t = 0$ ms denotes when the container begins to contact the substrate.
The time when the container separates from the substrate is defined as $t = \tau_{\rm impact}$.
In this section, $\tau_{\rm impact}$ is interpreted as the time when the impact interval ends, and $\tau_{\rm focusing}$ as the time when the focusing interval ends.
Here, we focus on the relative magnitudes of $\tau_{\rm impact}$ and $\tau_{\rm focusing}$.
In the \emph{rigid-impact}  regime, for $E = 2.0 \times 10^{5}$ MPa, $\tau_{\rm impact} = 0.2$ ms and $\tau_{\rm focusing} = 1.9$ ms, giving $\tau_{\rm impact} \ll \tau_{\rm focusing}$ (see figure \ref{fig:ejection}(a)).
This means that the impact interval is shorter than the focusing interval, and the jet forms while contact is still ongoing.
In contrast, in the \emph{soft-impact}  regime, the order of $\tau_{\rm impact}$ and $\tau_{\rm focusing}$ is reversed.
For $E = 1.7 \times 10^{0}$ MPa, $\tau_{\rm impact} = 6.5$ ms and $\tau_{\rm focusing} = 5.0$ ms.
For $E = 8.1 \times 10^{-1}$ MPa, $\tau_{\rm impact} = 9.2$ ms and $\tau_{\rm focusing} = 5.7$ ms, and in both cases $\tau_{\rm impact} > \tau_{\rm focusing}$.
This means that the impact interval is longer than the focusing interval, and the jet is formed during the contact (see figures \ref{fig:ejection}(b, c)).
The key point is that as $E$ decreases, the relationship between $\tau_{\rm impact}$ and $\tau_{\rm focusing}$ is reversed, and the impact interval becomes longer than the focusing interval.
This suggests that, in the \emph{soft-impact}  regime, the \emph{total impulse} from the substrate to the container is not fully transmitted to the concave gas--liquid interface before the jet is formed.

\subsubsection{Development of a new model equation for the jet velocity}\label{subsec:new model}

To discuss the mechanism behind the decrease in $V_{\rm j}/V_{\rm i}$ on soft substrates, we model the jet velocity in terms of the impulse transmitted from the substrate during the time that is dynamically relevant for formation.
In Pokrovski's experiment, the impact generates a force on the container over a finite duration, which initiates the motion of the concave gas--liquid interface and is subsequently amplified by flow focusing \citep{antkowiakShorttermDynamicsDensity2007, kiyama2014}.
In the \emph{rigid-impact}  regime, the contact ends well before the jet forms, and it is therefore natural to relate the jet to the \emph{total impulse} delivered during the contact.
In the \emph{soft-impact}  regime, however, we observe $\tau_{\rm impact} > \tau_{\rm focusing}$ (see figures \ref{fig:ejection}), meaning that the jet forms while contact is still ongoing.
In this case, the interface cannot be driven by the \emph{total impulse} accumulated over the entire contact duration; instead, jet formation can only depend on the impulse delivered up to the end of the focusing interval.

\begin{figure}
  \centerline{\includegraphics[width=1\textwidth]{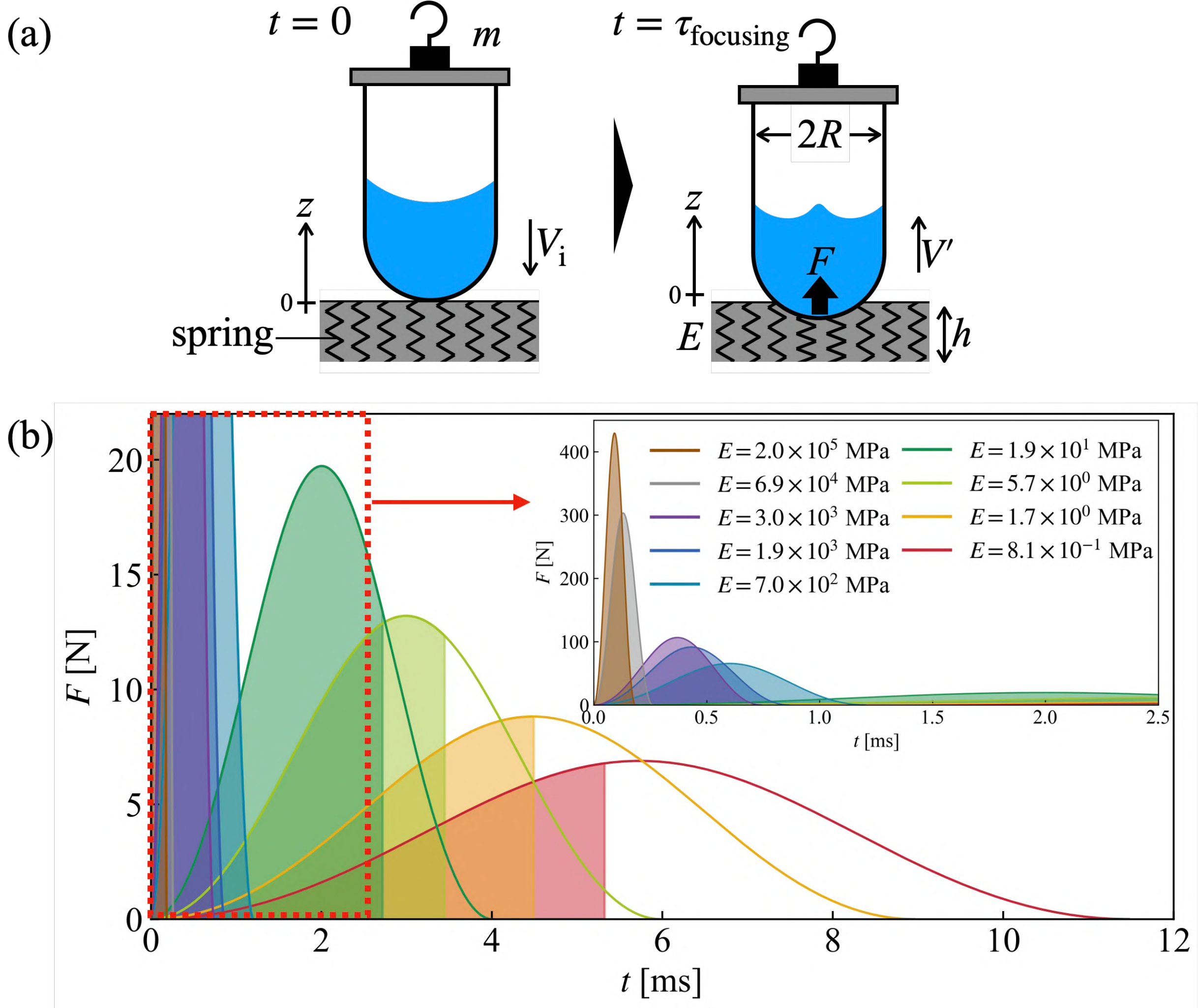}}
  \captionsetup{style=capcenter}
  \caption{(a) A schematic of the impactor and substrate at $t = 0$ and $t = \tau_{\rm focusing}$. $V'$ denotes the velocity of the impactor at $t = \tau_{\rm focusing}$.
In this study, $F$ is evaluated using an elastic foundation model, in which the elastic response of the substrate is modelled as a spring.
 (b) The time evolution of $F$ for each $E$, obtained from the elastic foundation model for $H = 20$ mm and $V_{\rm i} = 0.63$ m/s.
The shaded area represents $I$ calculated from equation (\ref{eq:Vj model}), i.e., the effective impulse time window. For $E \ge 5.7 \times 10^{0}$ MPa, which falls within the \emph{rigid-impact}  regime, the jet is driven by $I$ obtained by integrating $F$ over its entire duration, i.e. the \emph{total impulse}.
In contrast, for $E \le 1.9 \times 10^{1}$ MPa, which belongs to the \emph{soft-impact}  regime, the jet is driven by the impulse $I$ obtained from a partial integration of $F$, corresponding to a \emph{partial impulse}.}
\label{fig:impulse}
\end{figure}

This idea can be expressed without introducing additional variables by integrating the contact force over a finite time window.
Importantly, this formulation automatically recovers the \emph{rigid-impact}  limit.
When $\tau_{\rm impact} < \tau_{\rm focusing}$, the contact force vanishes for $t > \tau_{\rm impact}$, and thus integrating up to $t=\tau_{\rm focusing}$ is equivalent to integrating over the full contact.
In contrast, when $\tau_{\rm impact} > \tau_{\rm focusing}$, the upper limit $t=\tau_{\rm focusing}$ cuts off the force history, so that only a portion of the \emph{total impulse} contributes to jet formation (see figure \ref{fig:impulse}(b) and figures \ref{fig:numerical} (rigid-1) and (soft-1)–(soft-4) ).
In the remainder of this section, we translate this concept into a predictive model by combining momentum conservation with a contact-force model based on the elastic foundation model.

We consider the motion of the impactor of mass $m$ over the interval from first contact, $t = 0$, to the end of the focusing interval, $t = \tau_{\rm focusing}$.
Figure \ref{fig:impulse}(a) shows a schematic of the impactor and the substrate over the time interval [0,\, $\tau_{\rm focusing}$].
Let $F(t)$ denote the magnitude of the normal contact force exerted by the substrate on the container.
Over $[0,\tau_{\rm focusing}]$, momentum conservation gives

\begin{equation}
m(V_{\mathrm{i}} + V') = \int_{0}^{\tau_{\rm focusing}} F(t)\: dt.
  \label{eq:momentum conservation}
\end{equation}

\noindent
Here, $V'$ is the velocity of the impactor at $t = \tau_{\rm focusing}$.
Equation (\ref{eq:momentum conservation}) therefore relates the impulse delivered up to $t=\tau_{\rm focusing}$ to the momentum change of the impactor over $[0,\tau_{\rm focusing}]$.

Next, we relate the left-hand side of equation (\ref{eq:momentum conservation}) to the jet velocity $V_{\rm j}$.
Based on pressure--impulse theory, \citet{kiyama2014} related the change in momentum of the impactor to $V_{\rm j}$.
That is, $V_{\rm j}$ can be described by considering the change in momentum of the impactor over [0, $\tau_{\rm focusing}$] as

\begin{equation}
V_{\rm j} = \alpha (V_{\mathrm{i}} + V').
 \label{eq:VjViV'}
\end{equation}

\noindent
$\alpha$ is the strength of the flow-focusing effect after the interface acquires velocity due to the impact \citep{kiyamaEffectsWaterHammer2016}.
The value of $\alpha$ depends on the interface geometry (e.g. the contact angle at the wall) and on viscous effects \citep{kiyama2014,chengViscousInfluencesImpulsively2024}.
In this study, the strength of the flow-focusing effect is calculated using the semi-empirical formula derived by \citet{chengViscousInfluencesImpulsively2024} from experiments and simulations on rigid substrates.
This semi-empirical formula accounts for the viscous boundary layer and is expressed as
\begin{equation}
\alpha = 2.05(1-e^{- \zeta Re_{\rm p}}).
  \label{eq:alpha}
\end{equation}

\noindent
$Re_{\rm p}$ denotes the Reynolds number and can be written as

\begin{equation}
Re_{\rm p} = \sqrt{\frac{R U_0}{\nu}}.
  \label{eq:Rep}
\end{equation}

\noindent
$U_0$ is the velocity of the concave gas--liquid interface just after the impact, $\nu$ is the kinematic viscosity, and $\zeta = 0.076$ is a fitting parameter \citep{chengViscousInfluencesImpulsively2024}.
For rigid substrates, $ U_0$ can be defined as $U_0 = V_{\rm i} + V_{\rm r}$, where $V_{\rm r}$ is the rebound velocity of the impactor \citep{kiyamaEffectsWaterHammer2016, kiyama2014, chengViscousInfluencesImpulsively2024}.
We obtained $U_0$ from the rigid substrate cases following the same procedure as in \citet{kiyamaEffectsWaterHammer2016, kiyama2014} and \citet{chengViscousInfluencesImpulsively2024}, which yields $\alpha = 1.91$ under the present experimental conditions.
Equations (\ref{eq:momentum conservation})--(\ref{eq:Rep}) therefore relate the jet velocity to the time integral of the contact force over $[0,\tau_{\rm focusing}]$.

We now model the right-hand side of equation (\ref{eq:momentum conservation}).
To describe the time history $F(t)$, we consider the one-dimensional contact dynamics of an impactor with a hemispherical bottom impacting a compliant substrate.
Let $z(t)$ denote the indentation of the substrate surface at the contact point (so that $z = 0$ at first contact and $z < 0$ during compression).
Since the gravitational term is negligibly small compared with the other terms, the equation of motion is

\begin{equation}
m \frac{d^2 z}{dt^2} + F(z) = 0.
  \label{eq:motion}
\end{equation}

\noindent
We apply the initial condition at $t = 0$, $dz/dt = -V_{\rm i}$ and $z = 0$, to equation (\ref{eq:motion}).
The contact force $F(z)$ can be theoretically obtained from the elastic foundation model using the displacement of the substrate surface \citep{johnsonContactMechanics1987}, which is appropriate for finite-thickness substrates.
For a rigid sphere (radius $R$) indenting a substrate of thickness $h$, the model gives

\begin{equation}
F(z) = -\frac{\pi R E}{h} z ^2.
 \label{eq:elastic foundation}
\end{equation}

\noindent
The force model is validated through comparison with our measurements.
The validity of the elastic foundation model for the present experiments is discussed in detail in Appendix \ref{app:Validity model}.
As shown in figure \ref{fig:impulse}(a), the elastic foundation model represents the substrate's elastic force using springs and describes the relationship between the contact force and the substrate surface displacement.
The time variation of the contact force $F(t)$ from the elastic foundation model can be obtained numerically using equations (\ref{eq:motion}) and (\ref{eq:elastic foundation}).
First, by substituting equation (\ref{eq:elastic foundation}) into equation (\ref{eq:motion}), we obtain a nonlinear second-order ordinary differential equation (ODE) in $z$ and $t$.
We solve this ODE numerically using an explicit Runge--Kutta method (the solve\_ivp function, an ODE solver from the SciPy library in Python) to obtain the time evolution of the substrate surface displacement $z = z(t)$.
Finally, by substituting the numerically obtained $z = z(t)$ into equation (\ref{eq:elastic foundation}), we can calculate $F(t)$.
As seen from equations (\ref{eq:motion}) and (\ref{eq:elastic foundation}), $F(t)$ is determined by $m$, $h$, $R$, $V_{\rm i}$ and $E$, but in the present experiments the most dominant parameter is $E$.

Figure \ref{fig:impulse}(b) shows the time evolution of $F$ for each $E$ as solid lines for $H = 20$ mm and $ V_{\rm i} = 0.63$ m/s.
The time when $F$ becomes positive as the substrate, and the container begins to make contact, is defined as $t = 0$ ms.
The end of the solid line indicates the time $t = \tau_{\rm impact}$, when $F$ returns to zero as the substrate and the container separate.
As $E$ decreases, the peak of $F$ decreases, while $\tau_{\rm impact}$ increases, consistent with softer contact.

Finally, we estimate $\tau_{\rm focusing}$ by assuming that a concave gas--liquid interface with a meniscus thickness $H_{\rm m}$ (see figure \ref{fig:ejection}) deforms at a velocity $ V_{\rm j}$, as

\begin{equation}
\tau_{\rm focusing} \approx \frac{H_{\rm m}}{V_{\rm j}}.
 \label{eq:tau}
\end{equation}

\noindent
As shown in figure \ref{fig:ejection}, $H_{\rm m}$ is defined as the distance between the upper and lower edges in the central region of the concave gas–liquid interface.
By measuring the meniscus thickness just before impact for all experimental conditions, we obtained $H_{\rm m} = 5.41$ mm.
Here, $H_{\rm m}$ can also be expressed using $R$.
Using the contact angle $\theta$ formed between the concave gas--liquid interface and the container wall and neglecting the wall thickness, $H_{\rm m}$ can be written geometrically as $H_{\rm m} = (R / \cos \theta) (1 - \sin \theta)$.

By using equation (\ref{eq:momentum conservation})--(\ref{eq:tau}), the impulse used to drive the concave gas--liquid interface $I$ can be modelled.
Therefore, a new closed model equation for the jet velocity can be written as

\begin{equation}
\frac{V_{\rm j}}{V_{\rm i}}
= \frac{\alpha }{m V_{\rm i}} \int_{0}^{{\frac{H_{\rm m}}{V_{\rm j}}}}  F(t) \: dt.
  \label{eq:Vj model}
\end{equation}

\noindent
Equation (\ref{eq:Vj model}) reduces to the \emph{rigid-impact}  limit when $H_{\rm m}/V_{\rm j} > \tau_{\rm impact}$, because the integral then covers the full force history.
In the \emph{soft-impact}  regime, where $H_{\rm m}/V_{\rm j} < \tau_{\rm impact}$, the upper limit $t=H_{\rm m}/V_{\rm j}$ cuts off the integral, capturing the \emph{partial impulse} mechanism proposed above.
Based on the above considerations, $I$ is formulated as the right-hand side of equation (\ref{eq:Vj model}) as

\begin{equation}
I= \frac{\alpha }{m V_{\rm i}} \int_{0}^{{\frac{H_{\rm m}}{V_{\rm j}}}}  F(t) \: dt.
  \label{eq:I}
\end{equation}

\subsubsection{Numerical solution of the model equation}\label{subsec:numerical solution}

We further discuss $I$ by numerically solving the model equation and comparing the results with the experimental data.
First, we describe the numerical procedure used to solve equation (\ref{eq:Vj model}).
Since $V_{\rm j}$ appears on both sides of equation (\ref{eq:Vj model}), the equation can be solved numerically by treating $V_{\rm j}$ as an unknown variable.
Specifically, the integration time window of $F$, which is determined by the relative magnitudes of $\tau_{\rm impact}$ and $H_{\rm m}/V_{\rm j}$, is chosen such that the left-hand and right-hand sides of equation (\ref{eq:Vj model}) are equal.
This procedure yields the numerical solution for $V_{\rm j}$.

\begin{figure}
  \centerline{\includegraphics[width=1\textwidth]{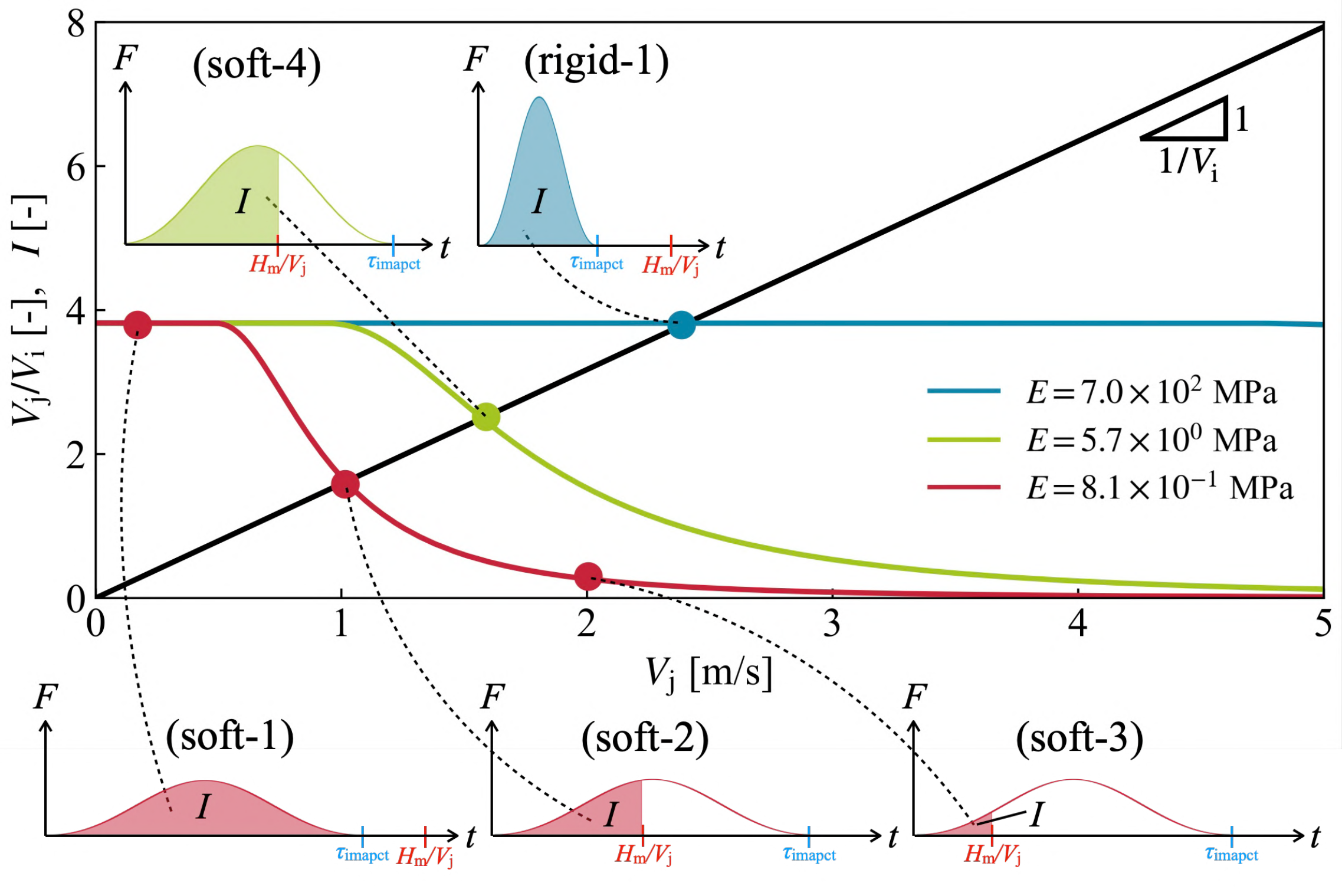}}
  \captionsetup{style=capcenter}
  \caption{The relationship between an arbitrary $V_{\rm j}$, the left-hand side of the equation (\ref{eq:Vj model}) $(V_{\rm j}/V_{\rm i})$, and the right-hand side $(I)$.
The black line represents $V_{\rm j}/V_{\rm i}$.
The red, green and blue lines indicate $I$ for $E = 8.1 \times 10^{-1}$ MPa, $E = 5.7 \times 10^{0}$ MPa and $E = 7.0 \times 10^{2}$ MPa, respectively.
For each value of $E$, the numerical solution for $V_{\rm j}$ is given by the intersection of the black line with the corresponding coloured line. 
(rigid-1) denotes the value of $I$ at the numerical solution for $V_{\rm j}$ in the case of $E = 7.0 \times 10^{2}$ MPa, which belongs to the \emph{rigid-impact}  regime. In this case, $I$ is obtained by integrating $F$ over its entire duration.
(soft-2) and (soft-4) correspond to the values of $I$ at the numerical solutions for $V_{\rm j}$ when $E = 8.1 \times 10^{-1}$ MPa and $E = 5.7 \times 10^{0}$ MPa, which belongs to the \emph{soft-impact}  regime, respectively. Here, the integration window of $F$ is limited to $t = H_{\rm m}/V_{\rm j}$, and $I$ is calculated from a partial integration of $F$.
(soft-1) to (soft-3) illustrate the relationship between $V_{\rm j}$ and $I$, together with the integration window, for $E = 8.1 \times 10^{-1}$ MPa.
These results demonstrate that $I$ represents either the \emph{total impulse} or a \emph{partial impulse} depending on the relative magnitudes of $H_{\rm m}/V_{\rm j}$ and $\tau_{\rm impact}$.}
\label{fig:numerical}
\end{figure}

First, we examine how the left-hand and right-hand sides of equation (\ref{eq:Vj model}) vary as functions of an arbitrary $V_{\rm j}$.
Figure \ref{fig:numerical} shows how both sides of equation (\ref{eq:Vj model}) depend on $V_{\rm j}$ for different values of $E$ with $H = 20$ mm and $V_{\rm i} = 0.63$ m/s.
The $x$-axis represents an arbitrary $V_{\rm j}$, while the $y$-axis shows $V_{\rm j}/V_{\rm i}$ (line $\text{—}$) and $I$ (lines \textcolor{red}{$\text{—}$}, \textcolor{YellowGreen}{$\text{—}$} and \textcolor{cyan}{$\text{—}$}).
The red, green and blue lines correspond to $E = 8.1 \times 10^{-1}$ MPa, $E = 5.7 \times 10^{0}$ MPa and $E = 7.0 \times 10^{2}$ MPa, respectively.
For each value of $E$, the numerical solution for $V_{\rm j}$ is given by the intersection of the black line ($\text{—}$) with the corresponding coloured line (\textcolor{red}{$\text{—}$}, \textcolor{YellowGreen}{$\text{—}$} and \textcolor{cyan}{$\text{—}$}).

First, the value of $V_{\rm j}/V_{\rm i}$ increases monotonically with a slope determined by $V_{\rm i}$.
In contrast, the behaviour of $I$ varies depending on both the force profile $F$, which is primarily governed by $E$, and the integration time window, which is determined by the relative magnitudes of $\tau_{\rm impact}$ and $H_{\rm m}/V_{\rm j}$, as illustrated by (rigid-1) and (soft-1)–(soft-4) in figure \ref{fig:numerical}.
For $E = 7.0 \times 10^{2}$ MPa, which falls within the \emph{rigid-impact}  regime, $I$ remains constant.
In contrast, for $E = 5.7 \times 10^{0}$ MPa and $E = 8.1 \times 10^{-1}$ MPa, both of which belong to the \emph{soft-impact}  regime, $I$ exhibits a constant followed by a decreasing region.

First, we consider the case of $E = 7.0 \times 10^{2}$ MPa.
As shown in figure \ref{fig:numerical} (rigid-1), for $ V_{\rm j}$ up to 5 m/s, $H_{\rm m}/V_{\rm j}$ is larger than $\tau_{\rm impact}$, and therefore the integration time window of $F$ is [0, $\tau_{\rm impact}$].
In this case, $F$ is integrated over its entire duration.
Because the integration time window does not depend on $V_{\rm j}$, the value of $I$ remains constant regardless of $V_{\rm j}$.
Figure \ref{fig:numerical} (rigid-1) shows the value of $I$ at which $V_{\rm j}$ becomes the numerical solution for $E = 7.0 \times 10^{2}$ MPa.
When $I$, obtained by integrating the full force history, equals $V_{\rm j}/V_{\rm i}$, the corresponding $V_{\rm j}$ is identified as the numerical solution.

Next, we consider the case of $E = 8.1 \times 10^{-1}$ MPa.
In the range up to approximately $V_{\rm j} = 0.6$ m/s, as shown in figure \ref{fig:numerical} (soft-1), $H_{\rm m}/V_{\rm j}$ is larger than $\tau_{\rm impact}$, and therefore, the integration time window of  $F$ is [0, $\tau_{\rm impact}$].
In this regime, the entire force history is integrated, and thus $I$ takes the same value as in the case of $E = 7.0 \times 10^{2}$ MPa discussed above.
However, for $V_{\rm j}$ exceeding approximately 0.6 m/s, as shown in figure \ref{fig:numerical} (soft-2) and (soft-3), $H_{\rm m}/V_{\rm j}$ becomes smaller than $\tau_{\rm impact}$, so that the integration time window of $F$ is reduced to [0, $H_{\rm m}/V_{\rm j}$].
Because $F$ is then only partially integrated, the resulting $I$ decreases.
Moreover, as seen by comparing figure \ref{fig:numerical} (soft-2) and (soft-3), the integration time window [0, $H_{\rm m}/V_{\rm j}$] shrinks with increasing $V_{\rm j}$, leading to a reduction in $I$ as $V_{\rm j}$ increases.
Figure \ref{fig:numerical} (soft-2) shows the value of $I$ at which $V_{\rm j}$ becomes the numerical solution for $E = 8.1 \times 10^{-1}$ MPa.
When $I$, obtained from the partial integration of $F$, equals $V_{\rm j}/V_{\rm i}$, the corresponding $V_{\rm j}$ is identified as the numerical solution.

Finally, by examining the values of $I$ at which $V_{\rm j}$ becomes the numerical solution for each value of $E$, as shown in figure \ref{fig:numerical} (rigid-1), (soft-2) and (soft-4), we find that $I$ represents either the \emph{total impulse} or a \emph{partial impulse} depending on the relative magnitudes of $ H_{\rm m}/V_{\rm j}$ and $\tau_{\rm impact}$.
Thus, by treating $V_{\rm j}$ as an unknown variable and determining the integration time window such that $V_{\rm j}/V_{\rm i}$ equals $I$ for each value of $E$, the numerical solution of equation (\ref{eq:Vj model}) can be obtained.

\subsubsection{Mechanism of jet velocity reduction}\label{subsec:mechanism}

We validate equation (\ref{eq:Vj model}) by comparing its numerical solutions with the experimental results.
The red solid line in figure \ref{fig:Vj & scaling}(b) represents the results obtained by numerically solving the equation (\ref{eq:Vj model}) using an explicit Runge--Kutta method combined with trapezoidal integration.
Equation (\ref{eq:Vj model}) captures both the small variation of $V_{\rm j}/V_{\rm i}$ in the \emph{rigid-impact}  regime and its decrease in the \emph{soft-impact}  regime.
This shows that $I$ is well modelled.
Here, we discuss the mechanism of the decrease in $V_{\rm j}/V_{\rm i}$ in the \emph{soft-impact}  regime by focusing on $I$, as calculated from equation (\ref{eq:Vj model}).
The shaded regions in figure \ref{fig:impulse}(b) represent $I$ for each $E$ obtained from equation (\ref{eq:Vj model}) with $H = 20$ mm and $V_{\rm i} = 0.63$ m/s.
In the \emph{rigid-impact}  regime ($E = 1.0 \times 10^{3}$ MPa to $2.0 \times 10^{5}$ MPa), the \emph{total impulse} transferred from the substrate to the container is used to drive the concave gas--liquid interface.
In contrast, in the \emph{soft-impact}  regime ($E = 8.1 \times 10^{-1}$ MPa to $1.9 \times 10^{1}$ MPa), only the \emph{partial impulse} is used to drive the concave gas--liquid interface compared with the \emph{rigid-impact}  regime.
In the \emph{soft-impact}  regime, the impact interval becomes longer than the focusing interval, so the effective impulse time window that accelerates the fluid into a jet becomes relatively short.
As a result, fluid motion is driven by only the \emph{partial impulse} compared with the \emph{rigid-impact}  regime, leading to a decrease in $V_{\rm j}/V_{\rm i}$.
This is the mechanism responsible for the reduction of $V_{\rm j}/V_{\rm i}$ in the \emph{soft-impact}  regime.

In this study, we have introduced a new framework in which only the portion of the \emph{total impulse} exerted by the substrate on the container that is actually transmitted to the fluid within the time window effective for jet formation is assumed to drive the jet.
This framework enables a unified description of impact-driven jet phenomena across a wide range of substrate stiffness, including both rigid and soft substrates, which have been difficult to treat within conventional pressure--impulse theories.

\section{Conclusion}
\label{sec:conclusion}

In this study, we have investigated the impact of a fluid-filled container on a soft substrate.
The aim of this study was to clarify the mechanism by which the impulse $I$ used to deform the concave gas--liquid interface is reduced on soft substrates compared with rigid substrates, by focusing on the velocity of the jet generated from the interface after impact.

We conducted drop experiments with a partially liquid-filled test tube while varying the substrate elastic modulus $E$ over a wide range from $8.1 \times 10^{-1}$ MPa to $2.0 \times 10^{5}$ MPa.
We found that the jet velocity $V_{\rm j}$ decreases as $E$ decreases (see figure \ref{fig:img & Vtip} and figure \ref{fig:Vj & scaling}(a)).
We also examined the relationship between the impact interval $\tau_{\rm impact}$, defined as the time during which the container and the substrate are in contact, and the focusing interval $\tau_{\rm focusing}$, defined as the time used for a jet to form from a concave gas--liquid interface.
Importantly, we found experimentally that the impact interval becomes longer than the focusing interval as $E$ decreases (see figure \ref{fig:ejection}).
This implies that, for smaller values of $E$, the jet forms while the container remains in contact with the substrate.

Using dimensional analysis, we examined the relationship between $V_{\rm j}$ and $E$ in detail.
We found that the dimensionless jet velocity $V_{\rm j}/V_{\rm i}$ can be organized by the Cauchy number $\rho_{\rm e} V_{\rm i}^2 / E$, which represents the ratio of the inertial force in the container--liquid system to the elastic force of the substrate.
When $V_{\rm j}/V_{\rm i}$ is plotted against $\rho_{\rm e} V_{\rm i}^2 / E$, the experimental data collapse onto a single curve (see figure \ref{fig:Vj & scaling}(b)).
This plot reveals two regimes: one in which $V_{\rm j}/V_{\rm i}$ is nearly constant and another in which $V_{\rm j}/V_{\rm i}$ decreases significantly.
The region where $V_{\rm j}/V_{\rm i}$ changes little corresponds to $\rho_{\rm e} V_{\rm i}^2 / E< 10^{-4}$, and this region can be explained using the experimental results of \citet{kiyamaEffectsWaterHammer2016} obtained with rigid substrates.
This indicates that, in the region $\rho_{\rm e} V_{\rm i}^2 / E< 10^{-4}$, the jet velocity can be modelled using the conventional pressure--impulse theory established through impact experiments on rigid substrates \citep{batchelorIntroductionFluidDynamics1967, antkowiakShorttermDynamicsDensity2007, kiyama2014, kiyamaEffectsWaterHammer2016}.
Therefore, we defined the region $\rho_{\rm e} V_{\rm i}^2 / E< 10^{-4}$ as the \emph{rigid-impact}  regime.
In contrast, the region where $V_{\rm j}/V_{\rm i}$ decreases significantly corresponds to $\rho_{\rm e} V_{\rm i}^2 / E > 10^{-4}$.
In this region, the jet velocity cannot be modelled using conventional pressure--impulse theory.
We experimentally showed for the first time that the effect of soft substrates becomes significant in fluid motion when $\rho_{\rm e} V_{\rm i}^2 / E > 10^{-4}$.
Therefore, we defined the region $\rho_{\rm e} V_{\rm i}^2 / E > 10^{-4}$ as the \emph{soft-impact}  regime and discussed the mechanism of jet velocity reduction in this regime.

The key observation for elucidating the mechanism underlying the reduction of $V_{\rm j}/V_{\rm i}$ is the relative ordering of two timescales: the impact interval $\tau_{\rm impact}$ and the focusing interval $\tau_{\rm focusing}$.
In the \emph{rigid-impact}  regime, $\tau_{\rm impact}< \tau_{\rm focusing}$ holds (see figure \ref{fig:ejection}(a)).
In this case, the contact force acts only over a short time, and the fluid subsequently focuses after the contact has ended.
In this limit, it is natural to relate jet formation to the \emph{total impulse} delivered during contact.
In contrast, in the \emph{soft-impact}  regime, we find $\tau_{\rm impact} > \tau_{\rm focusing}$, indicating that jet formation starts while contact is still ongoing (see figures \ref{fig:ejection}(b,c)).
In this case, the jet cannot be driven by the \emph{total impulse} transmitted over the entire contact duration, because the jet forms before the \emph{total impulse} has been delivered to the fluid.
This observation motivates a framework in which the jet is governed by the impulse accumulated up to the jet-formation time, i.e., the \emph{partial impulse} (see figure \ref{fig:impulse}(b) and figures \ref{fig:numerical} (soft-2)--(soft-4)).

This framework can be expressed by integrating the contact force over a finite time window.
Importantly, the present formulation automatically recovers the \emph{rigid-impact} limit.
When $\tau_{\rm impact} < \tau_{\rm focusing}$, the contact force vanishes for $t > \tau_{\rm impact}$, and therefore, integrating up to $t=\tau_{\rm focusing}$ is equivalent to integrating over the entire contact duration.
In contrast, when $\tau_{\rm impact} > \tau_{\rm focusing}$, the upper limit $t=\tau_{\rm focusing}$ truncates the force history, so that only a portion of the \emph{total impulse} contributes to jet formation.
In this paper, we formulated this framework as a predictive model by combining momentum conservation with a contact force model based on the elastic foundation model.
As a result, the impulse $I$ was formulated as shown in equations (\ref{eq:Vj model}) and (\ref{eq:I}).
By comparing the experimental results with numerical solutions of the model equation, treating $V_{\rm j}$ as an unknown, we demonstrated that the model quantitatively accounts for the experimental observations (see figure \ref{fig:Vj & scaling}(b)).

The key contribution of this study is the introduction of a framework in which only the portion of the \emph{total impulse} transmitted from the substrate to the container that is actually delivered to the fluid within the effective time window for jet formation drives the jet.
This framework enables a unified description of impact-driven jetting over a wide range of substrate stiffness, including both rigid and soft substrates.
In particular, for sufficiently soft substrates, the contact duration between the substrate and the container becomes longer than the time required for jet formation, so that the effective time window during which the fluid is accelerated into a jet becomes relatively short.
As a result, in the case of soft substrates, the fluid motion is driven by only a \emph{partial impulse} compared with that for rigid substrates, leading to a reduction in the jet velocity.

\begin{appen}
\section{Method for measuring the substrate elastic modulus}
\label{app:measuring E}

In this study, the elastic modulus $E$ of rubber and elastomer substrates was measured experimentally.
In this section, we describe the method used to measure $E$.
The substrate elasticity can be estimated from the surface deformation $\delta$ when a rigid sphere is pressed vertically onto the substrate with a load $P$ (see figure \ref{fig:app_Eexp}(a)).
In the experiment, the substrate elastic modulus was calculated using the elastic foundation model \citep{johnsonContactMechanics1987}  as

\begin{equation}
\label{eq:exp elastic foundation}
E = \frac{Ph}{\pi R_{\mathrm{s}} \delta^2}.
\end{equation} 

\noindent
Here, $h$ denotes the substrate thickness, and $R_{\rm s}$ is the radius of the rigid sphere.
In this experiment, a steel sphere with a diameter of $2R_{\rm s} = 12.7$ mm was used.

Next, the experimental procedure is described.
As shown in figure \ref{fig:app_Eexp}(b), a metal rod was fixed with screws to a micrometre (Newport, M-562 Series) that can move vertically.
The steel sphere was fixed to a metal rod using adhesive.
An electronic balance (AS ONE Corporation, AXA20002) was placed beneath the steel sphere, and an aluminium plate and the substrate to be measured were placed on the balance.
The balance was zeroed with the aluminium plate and the substrate on it.
By moving the micrometre, the steel sphere moved vertically together with the micrometre (see figure \ref{fig:app_Eexp}(b)).
The micrometre was adjusted so that the distance between the steel sphere and the substrate was zero ($\delta = 0$).
At this position, the reading of the balance was zero.
From this state, the micrometre was moved downward by a displacement $\delta$, causing the steel sphere to move downward by $\delta$ and deform the substrate surface by $\delta$ (see figure \ref{fig:app_Eexp}(a)).
The applied load $P$ on the substrate (see figure \ref{fig:app_Eexp}(a)) was calculated as $P = mg$, where $m$ is the apparent mass measured by the balance and $g = 9.81$ m/s$^2$.
$\delta$ ranged from 10 $\mu$m to 430 $\mu$m in steps of 10 $\mu$m, and the corresponding $P$ was recorded for each $\delta$.
The substrate elastic modulus at each $\delta$ was then calculated using equation (\ref{eq:exp elastic foundation}).
Each measurement was repeated five times for each value of $\delta$.

Figure \ref{fig:app_EvsDelta} shows the relationship between the elastic modulus calculated from equation (\ref{eq:exp elastic foundation}) and $\delta$ for rubber and elastomer substrates.
Panels (a)--(d) correspond to rubber, PDMS10, PDMS20 and PDMS30, respectively.
The plotted values are the averages of five measurements, and the error bars represent the standard deviation.
The elastic modulus converges as $\delta$ increases.
In this study, the elastic modulus averaged over $\delta \geq$ 200 $\mu$m was adopted as the representative value $E$ for each substrate (the red lines in figure \ref{fig:app_EvsDelta} indicate the averaged value of the elastic modulus for $\delta \geq$ 200 $\mu$m corresponding to $E$).
Using this method, the values of $E$ for the rubber and elastomer substrates are summarized in table \ref{table}.

\begin{figure}
  \centerline{\includegraphics[width=1\textwidth]{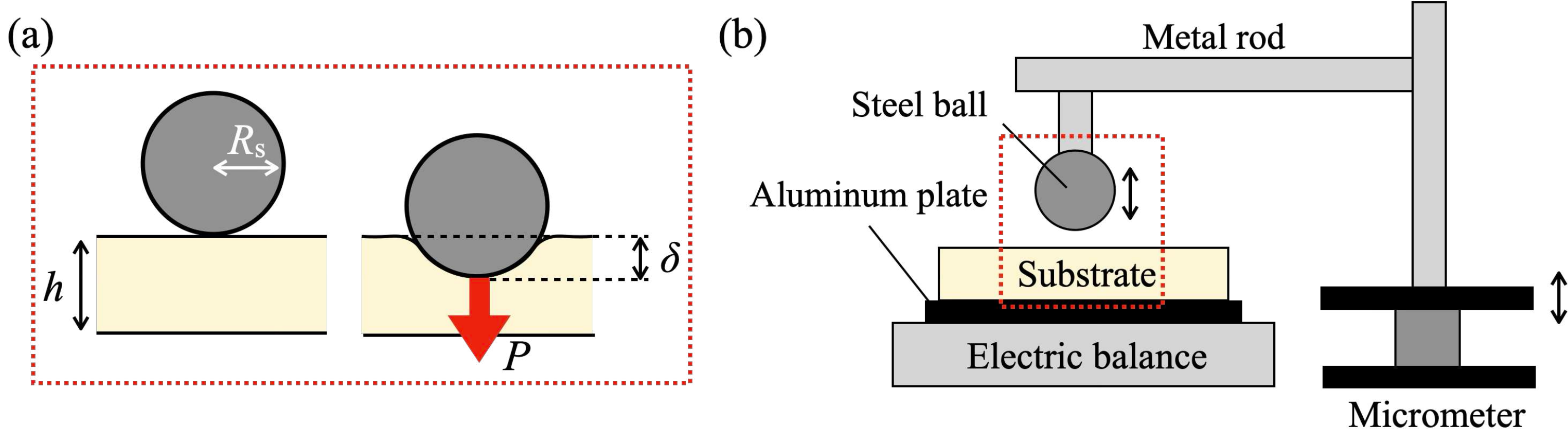}}
  \captionsetup{style=capcenter}
  \caption{(a) When $\delta$ is applied to the substrate surface, $P$ acts on the substrate. $h$ is the substrate thickness, and $R_{\rm s}$ is the radius of the iron ball. (b) A schematic of the experimental setup for measuring substrate elastic modulus.}
\label{fig:app_Eexp}
\end{figure}

\begin{figure}
  \centerline{\includegraphics[width=1\textwidth]{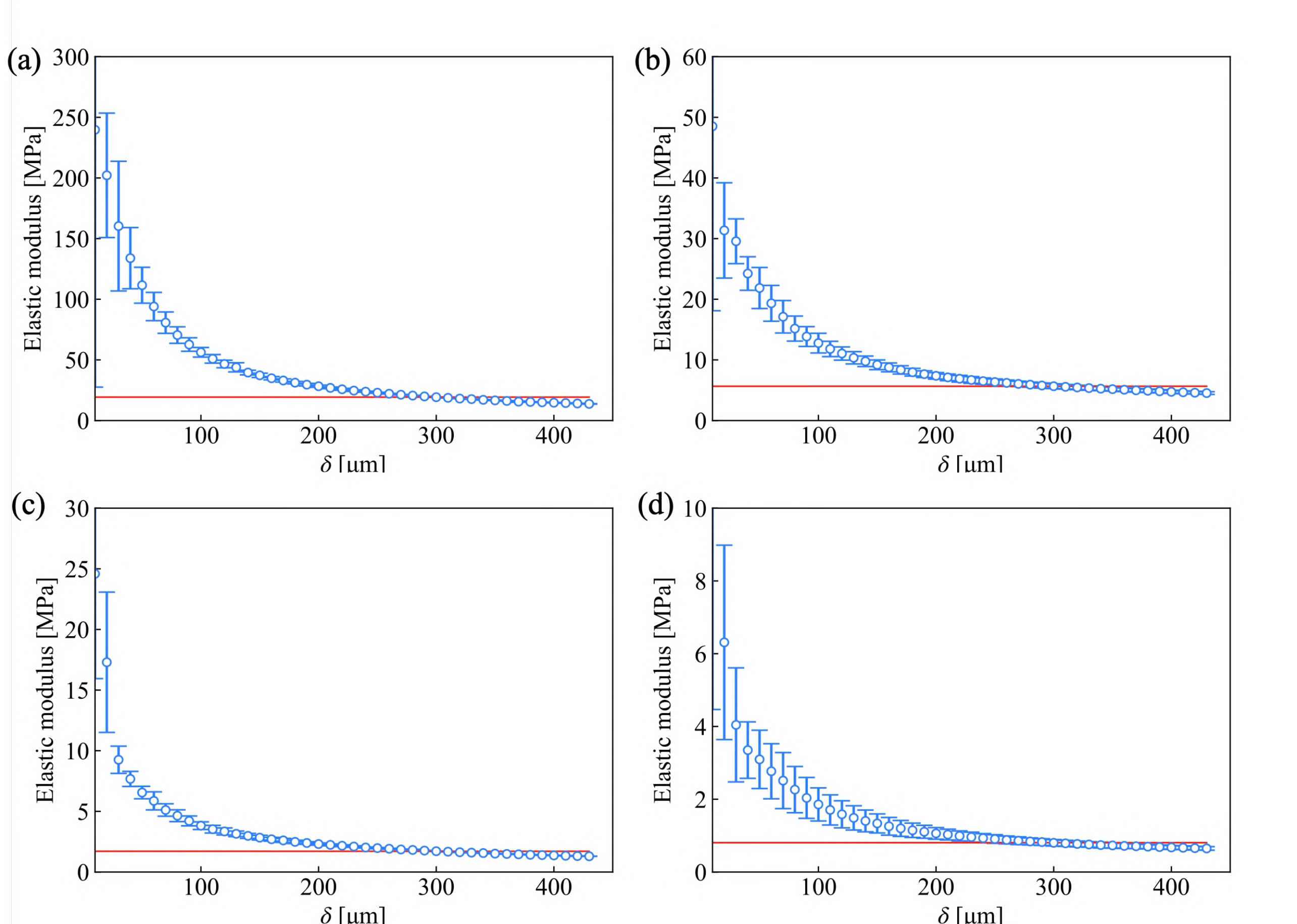}}
  \captionsetup{style=capcenter}
  \caption{The relationship between the elastic modulus calculated from equation (\ref{eq:exp elastic foundation}) and $\delta$ for (a) rubber, (b) PDMS10, (c) PDMS20, and (d) PDMS30. The red solid line represents the average elastic modulus for cases with $\delta \geq 200$ {\textmu}m corresponding to $E$.}
\label{fig:app_EvsDelta}
\end{figure}

\section{Fabrication procedure of the substrate using polydimethylsiloxane (PDMS)}
\label{app:PDMS}

In the experiment, polydimethylsiloxane (PDMS) elastomer substrates were prepared.
In this section, the fabrication procedure for the PDMS substrates is described.
First, the base material (DOW, SILPOT 184 Silicone Elastomer Base) and the curing agent (DOW, SILPOT 184 Silicone Elastomer Curing Agent) were thoroughly mixed at mass ratios of 10:1, 20:1 and 30:1.
Next, the mixed PDMS was degassed in a vacuum chamber for one hour.
The degassed PDMS was then poured into a container ($100 \times 100 \times 25$ mm) fabricated using a 3D printer (Bambu Lab A1 3D Printer) and degassed again for one hour.
During this process, care was taken to keep the container level to ensure a uniform substrate thickness.
The substrate was then heated at 80${}^\circ$C for two hours using a heater (Formlabs, Formlabs Form Cure L).
After heating and confirming that the PDMS was fully cured, the PDMS substrate was removed from the container.
Finally, baby powder (SiCCAROL-Hi, Asahi Group Foods Co., Ltd.) was applied to the PDMS substrate surface to remove surface tackiness.

\section{Validation of the elastic foundation model}
\label{app:Validity model}

In \S \ref{sec:mechanism}, a new model for the jet velocity was developed by modelling the impulse used to drive the concave gas--liquid interface based on the elastic foundation model.
In this section, the validity of the elastic foundation model is assessed by comparing the experimentally measured contact force with that calculated using the Hertz model.

\subsection{Measurement of the contact force using an accelerometer}

In this section, the method used to measure the contact force is described.
The experimental setup was based on the method proposed by \citet{kuriharaPressureFluctuationsLiquids2025}, and is shown in figure \ref{fig:app_Ac-exp}.
An accelerometer (Showa Sokki Co., Ltd., MODEL 2350, sensitivity 0.347 pC/(m/s$^2$)) was firmly fixed to an aluminium plate using screws, and the aluminium plate was then bonded to the top of the test tube.
A steel S-shaped hook with a jig was attached to the top of the accelerometer using double-sided tape, and the impactor was placed above the substrate using an electromagnet.
The accelerometer output an electric charge proportional to the applied acceleration.
The accelerometer was connected to a charge amplifier (Showa Sokki Co., Ltd., MODEL 4035-50) to convert the charge signal into a voltage signal.
The charge amplifier was connected to an oscilloscope (Iwasaki Communication Co., Ltd., DS-5554A), and the voltage time history was recorded.
The experimental conditions were the same as those described in \S \ref{sec:experimental setup}.

The raw voltage data recorded by the oscilloscope contained significant high-frequency noise, which was removed using a fast Fourier transform.
The filtered voltage signal [V] was then converted into acceleration $a$ [m/s$^2$] using the acceleration scale set in the charge amplifier.
The acceleration was further converted into force $F$ [N] using the equation of motion, $F = ma$.
In this experiment, $F$ was calculated by multiplying $a$ by the mass $m$ of the impactor.

The black solid lines in figure \ref{fig:app_Elastic_hertz_exp} show the time histories of $F$ measured using the accelerometer ($H = 20$ mm and $V_{\rm i} = 0.63$ m/s).
Panels (a)--(d) correspond to $E = 2.0 \times 10^{5}$ MPa, $E = 1.9 \times 10^{3}$ MPa, $E = 5.7 \times 10^{0}$ MPa and $E = 8.1 \times 10^{-1}$ MPa, respectively.
In panels (b)--(d), $F$ rises rapidly, reaches a peak, and then decreases.
In contrast, panel (a) shows two peak values.
The second peak is likely caused by the accelerometer detecting pressure fluctuations in the liquid \citep{kuriharaPressureFluctuationsLiquids2025}.

\begin{figure}
  \centerline{\includegraphics[width=0.6\textwidth]{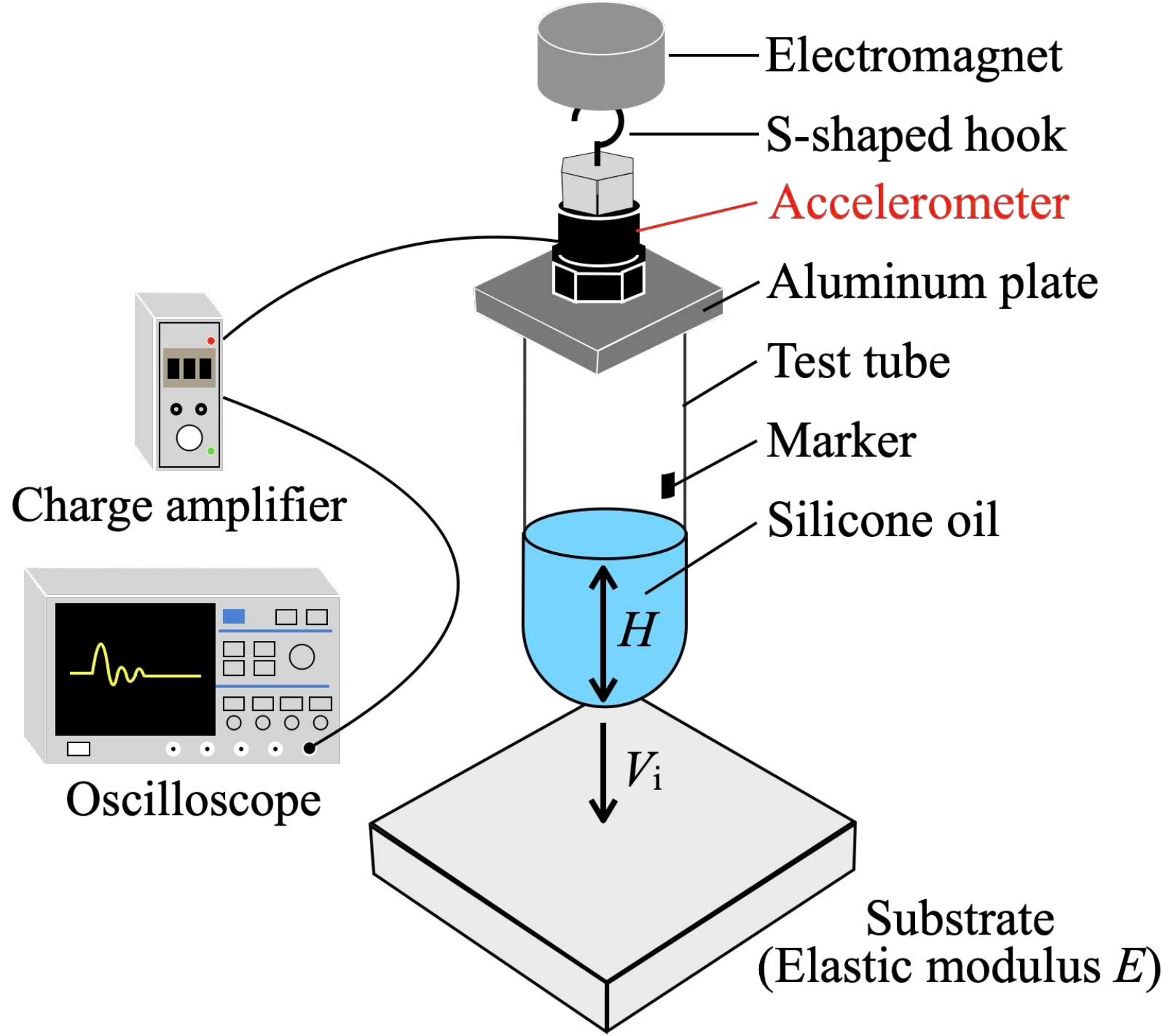}}
  \captionsetup{style=capcenter}
  \caption{A schematic of the experimental setup for measuring the contact force using an accelerometer.}
\label{fig:app_Ac exp}
\end{figure}

\begin{figure}
  \centerline{\includegraphics[width=1\textwidth]{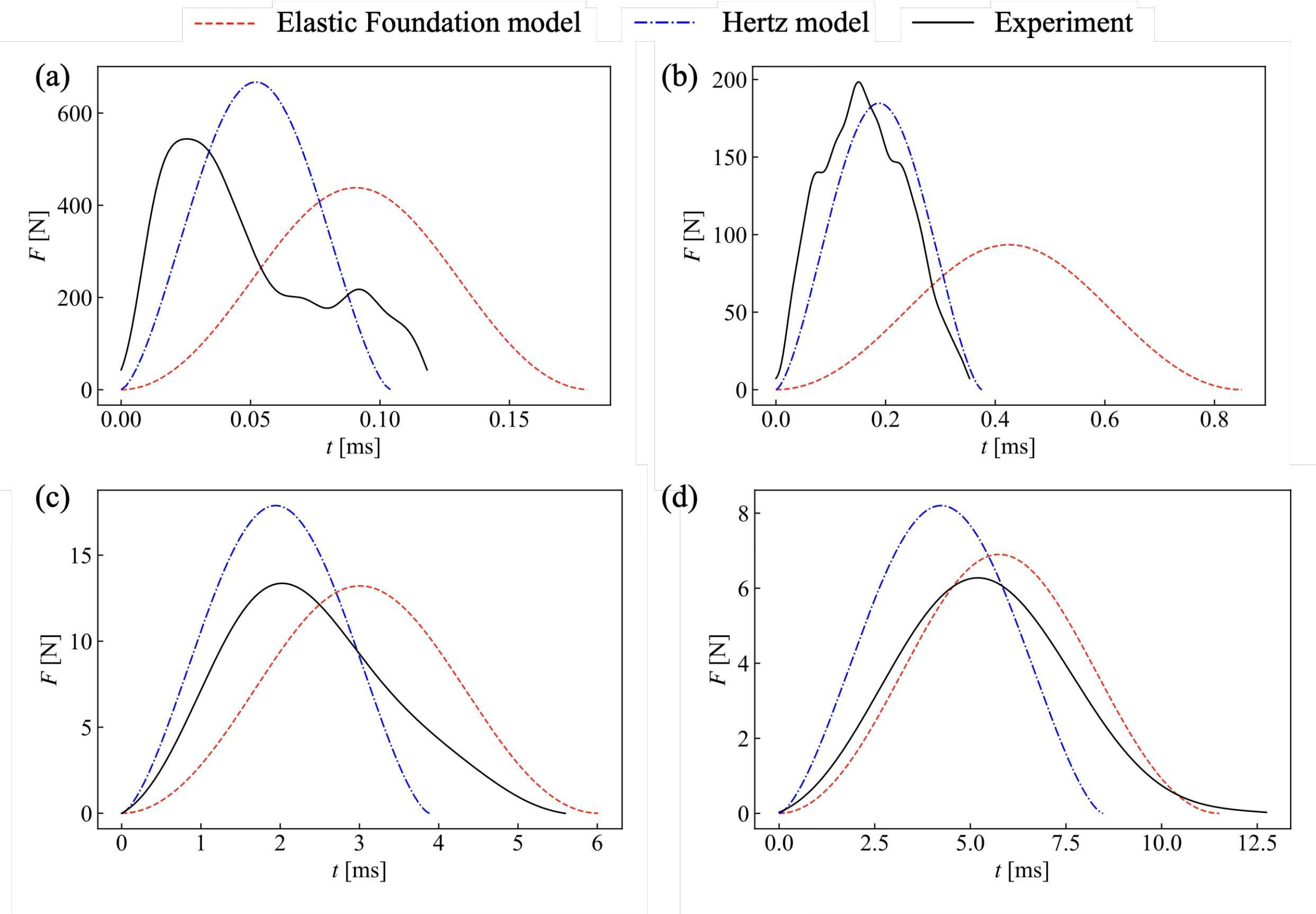}}
  \captionsetup{style=capcenter}
  \caption{A comparison of the experimental force $F$ with the Hertz model and the elastic foundation model. Panels (a)--(d) correspond to $E = 2.0 \times 10^{5}$ MPa, $E=1.9 \times 10^{3}$ MPa, $E=5.7 \times 10^{0}$ MPa, and $E = 8.1 \times 10^{-1}$ MPa, respectively. The conditions are $H = 20$ mm and $V_{\rm i} = 0.63$ m/s.}
\label{fig:app_Elastic_hertz_exp}
\end{figure}

\subsection{Comparison between the experimental contact force and impact models}

In \S \ref{sec:mechanism}, the impact of an object with a hemispherical bottom onto the substrate was modelled as the impact of a rigid sphere.
In \S \ref{sec:mechanism}, the elastic foundation model was used to model the contact force. However, the Hertz model is another representative model for the impact of a rigid sphere onto a substrate.
In this section, the differences between the elastic foundation model and the Hertz model are discussed.
In addition, the contact force measured with the accelerometer is compared with predictions from the elastic foundation and Hertz models to assess the validity of using the elastic foundation model.

The main difference between the elastic foundation model and the Hertz model lies in the potential of the contact displacement.
Figure \ref{fig:app_hertz_elastic} shows the displacement potentials for the two impact models.
The Hertz model assumes contact between a rigid sphere and a substrate with infinite thickness, and also assumes that the surface deformation of the substrate is very small \citep{johnsonContactMechanics1987}.
In this case, the contact displacement follows a spherical potential (see figure \ref{fig:app_hertz_elastic}).
Therefore, the Hertz model is suitable for modelling impacts between stiff substrates and rigid spheres.
However, the Hertz model cannot fully capture real behaviour when the substrate is thin or when the surface deformation is large compared to the substrate thickness.
In contrast, the elastic foundation model assumes contact between a rigid sphere and a substrate with finite thickness \citep{johnsonContactMechanics1987}, for which the contact displacement follows a vertical potential (see figure \ref{fig:app_hertz_elastic}).
Therefore, the elastic foundation model is effective for modelling impacts in cases where the effect of substrate thickness cannot be neglected.

In the Hertz model, $F$ is expressed in terms of the effective elastic modulus $E^*$, the radius of the rigid sphere $R$ and the surface displacement of the substrate $z$, as 

\begin{equation}
\label{eq:Hertz}
F = \frac{4}{3} E^* R ^{1/2} z ^{3/2}.
\end{equation}

\noindent
$E^*$ is given in terms of the elastic modulus $E_1$ and $E_2$ and the Poisson's ratios $\nu_1$ and $\nu_2$ of the rigid sphere and the substrate, respectively, as

\begin{equation}
\label{eq:E*}
E^* = \Big(\frac{1-\nu_1^2}{E_1} + \frac{1-\nu_2^2}{E_2} \Big) ^{-1}.
\end{equation}

\noindent
To calculate the contact force acting on the container in this experiment using the Hertz model, the container can be treated as a rigid sphere with a hemispherical bottom.
Specifically, the contact force predicted by the Hertz model can be obtained by numerically calculating equation (\ref{eq:motion}), substituting in equations (\ref{eq:Hertz}) and (\ref{eq:E*}).

Next, the contact force measured in the experiment is compared with the predictions of the elastic foundation model and the Hertz model.
Figure \ref{fig:app_Elastic_hertz_exp} shows a comparison for $F$.
The red dashed lines represent the elastic foundation model, and the blue dashed lines represent the Hertz model.
Here, we focus mainly on the end time of the $F$, which corresponds to the moment when the container separates from the substrate, when comparing the two models with the experimental data.
In the \emph{rigid-impact}  regime shown in figures \ref{fig:app_Elastic_hertz_exp}(a, b), the force profile predicted by the Hertz model captures the experimental trend better.
In contrast, in the \emph{soft-impact}  regime shown in figures \ref{fig:app_Elastic_hertz_exp}(c,d), the elastic foundation model reproduces the experimental trend more accurately.
These results indicate that the impact model that best describes the experimental force profile depends on $E$.

Based on this observation, we consider which impact model is appropriate for the present study.
We argue that, for the \emph{soft-impact}  regime, the impact interval becomes longer than the focusing interval, resulting in a relatively shorter effective impulse time window that accelerates the fluid into a jet.
From this perspective, the most appropriate impact model is the one that accurately captures the end time of $F$ at small $E$.
Therefore, the elastic foundation model is adopted in this study.

\begin{figure}
  \centerline{\includegraphics[width=0.6\textwidth]{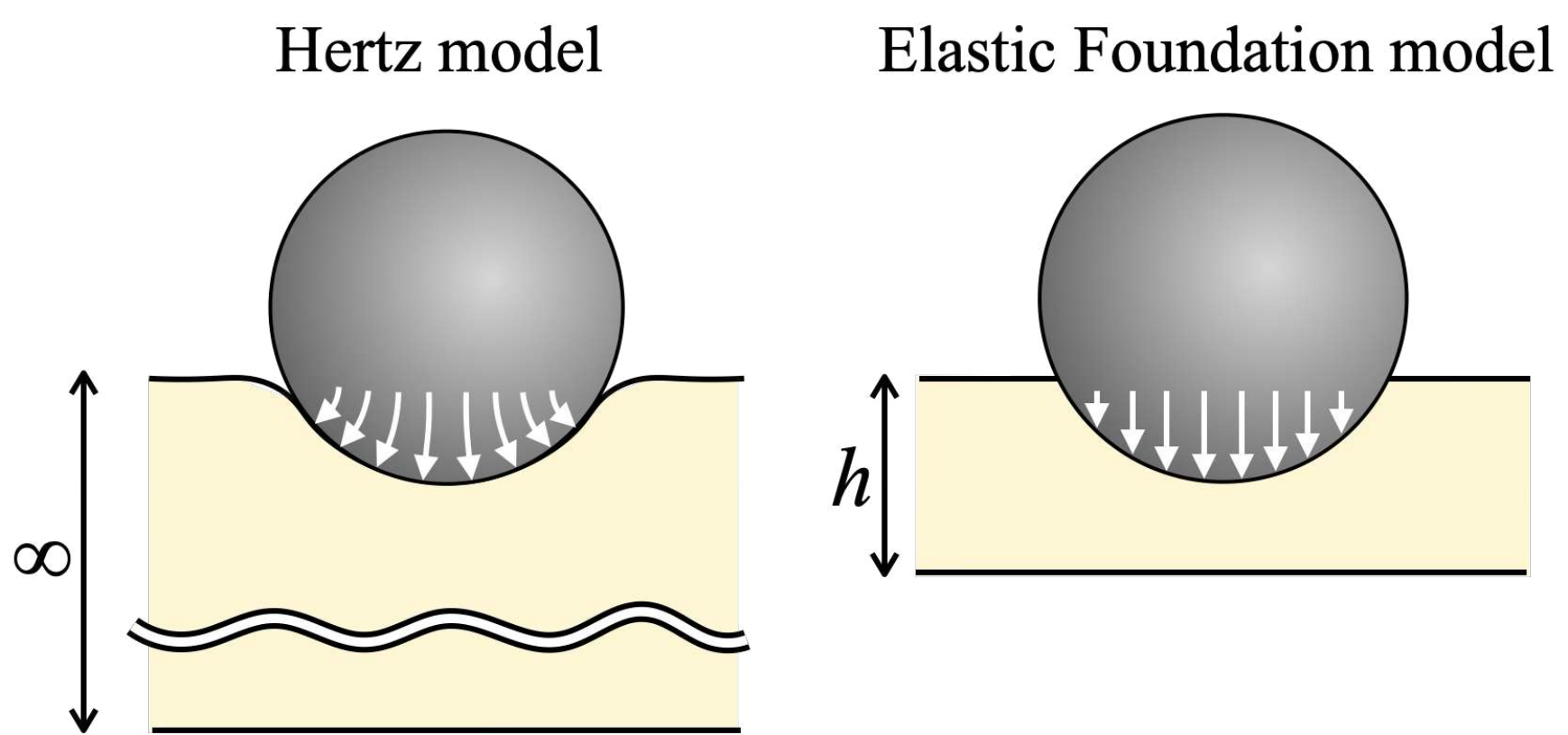}}
  \captionsetup{style=capcenter}
  \caption{A comparison of contact-displacement potentials between the Hertz model and the elastic foundation model.}
\label{fig:app_hertz_elastic}
\end{figure}




\end{appen}
\clearpage

\begin{bmhead}[Acknowledgments]
{This work was funded by the Japan Society for the Promotion of Science (Grant Nos. 20H00222, 20H00223, 20K20972, and 24H00289) and the Japan Science and Technology Agency (Grant Nos. PRESTO JPMJPR21O5 and SBIR JPMJST2355).}
\end{bmhead}

\begin{bmhead}[Declaration of Interests]
{The authors report no conflict of interest.}
\end{bmhead}

\bibliographystyle{jfm}
\bibliography{jfm}

\end{document}